%% file: HQL06_proc.tex
\documentclass[12pt]{article}
\usepackage{epsfig}

\input econfmacros.tex

\def\mevc  {\ifmmode {\rm MeV}/c \else MeV$/c$\fi}
\def\mevcc {\ifmmode {\rm MeV}/c^2 \else MeV$/c^2$\fi}
\def\gevc  {\ifmmode {\rm GeV}/c \else GeV$/c$\fi}
\def\gevcc {\ifmmode {\rm GeV}/c^2 \else GeV$/c^2$\fi}
\def\ra    {\rightarrow}
\def\Bs    {\ensuremath{B_s^0}}
\def\Bd    {\ensuremath{B^0}}
\def\Sb    {\ensuremath{\Sigma_b}}
\def\Bhh   {\ensuremath{B\ra h^+h^-}}
\def\Lb    {\ensuremath{\Lambda_b^0}}
\def\Lc    {\ensuremath{\Lambda_c^+}}
\def\Sib   {\ensuremath{\Sigma_b}}
\def\dedx  {\ifmmode {\rm d}E/{\rm d}x \else d$E/$d$x$\fi}
\newcommand{\Bdpipi}{\ensuremath{\Bd \to \pi^+ \pi^-}}
\newcommand{\BdKpi}{\ensuremath{\Bd \to K^+ \pi^-}}
\newcommand{\BsKpi}{\ensuremath{\Bs \to K^- \pi^+}}
\newcommand{\Bspipi}{\ensuremath{\Bs \to  \pi^+ \pi^-}}
\newcommand{\BsKK}{\ensuremath{\Bs \to  K^+ K^-}}
\newcommand{\BdKK}{\ensuremath{\Bd \to  K^+ K^-}}
\newcommand{\DKpi}{\ensuremath{D^{0} \to K^- \pi^+}}
\newcommand{\Lbppi}{\ensuremath{\Lambda_{b}^{0} \to p\pi^{-}}}
\newcommand{\LbpK}{\ensuremath{\Lambda_{b}^{0} \to pK^{-}}}
\newcommand{\aBdKpi}{\ensuremath{\bar{B}^0 \to K^- \pi^+}}
\newcommand{\aBsKpi}{\ensuremath{\bar{B}_s^0\to K^+ \pi^-}}
\newcommand{\acpbdkpi}{\ensuremath{{\cal A}_{CP}(\BdKpi)}}
\newcommand{\acpbskpi}{\ensuremath{{\cal A}_{CP}(\BsKpi)}}

\newcommand{\acpDKpi}{\ensuremath{{\cal A}_{CP}(\DKpi)}}
\newcommand{\br}{\ensuremath{\mathcal B}}
\newcommand{\ACPddef}{\ensuremath{{\frac{\br (\aBdKpi)-\br 
(\BdKpi)}{\br (\aBdKpi)+\br (\BdKpi)}}}}
\newcommand{\ACPsdef}{\ensuremath{{\frac{\br (\aBsKpi)-\br 
(\BsKpi)}{\br (\aBsKpi)+\br (\BsKpi)}}}}
\newcommand{\BdpipisuBdKpidef}{\ensuremath{\frac{\br(\Bdpipi)}{\br(\BdKpi)}}}
\newcommand{\BsKKsuBdKpidef}{\ensuremath{\frac{\mathit{f_s}}{\mathit{f_d}}\frac{\br(\BsKK)}{\br(\BdKpi)}}}

\newcommand{\BsKpisuBdKpidef}{\ensuremath{\frac{\mathit{f_s}}{\mathit{f_d}}\frac{\br(\BsKpi)}{\br(\BdKpi)}}}
\newcommand{\BspipisuBdKpidef}{\ensuremath{\frac{\mathit{f_s}}{\mathit{f_d}}\frac{\br(\Bspipi)}{\br(\BdKpi)}}}
\newcommand{\BdKKsuBdKpidef}{\ensuremath{\frac{\br(\BdKK)}{\br(\BdKpi)}}}

\newcommand{\LbppisuLbpK}{\ensuremath{\mathcal{B}(\Lbppi)/\mathcal{B}(\LbpK)}}

\newcommand{\LbppisuLbpKdef}{\ensuremath{\frac{\br(\Lbppi)}{\br(\LbpK)}}}

\newcommand{\rateratiodef}{\ensuremath{\frac{\mathit{f_d}}{\mathit{f_s}}{\frac{ \Gamma(\aBdKpi)- 
\Gamma(\BdKpi)}{\Gamma(\aBsKpi)-\Gamma(\BsKpi)}}}}

\def\Title#1{\begin{center} {\Large {\bf #1} } \end{center}}

\begin{document}

\Title{Spectroscopy and Decay of \boldmath{$B$}~Hadrons at the Tevatron}

\begin{center}{\large \bf Contribution to the proceedings of HQL06,\\
Munich, October 16th-20th 2006}\end{center}

\bigskip\bigskip


\begin{raggedright}  

{\it Manfred Paulini\index{Paulini, M.}\\
for the CDF and D\O~Collaboration \\
Carnegie Mellon University\\
Department of Physics\\
Pittsburgh, PA 15213, U.S.A.}
\bigskip\bigskip
\end{raggedright}

\section{Introduction}

Traditionally, $B$~physics has been the domain of $e^+ e^-$ machines
operating on the $\Upsilon(4S)$ resonance or the $Z^0$ pole.  But the
UA\,1~Collaboration has already shown that $B$~physics is feasible at a
hadron collider environment (see for example Ref.~\cite{bfeasi}).  The
first signal of fully reconstructed $B$~mesons at a hadron collider has
been published by the CDF~Collaboration in 1992~\cite{cdf_firstB}.  CDF
reconstructed a handful of $B^+ \ra J/\psi K^+$ events in a data sample of
2.6~pb$^{-1}$ taken during the Tevatron Run\,0 at the end of the
1980's. Since then experimental techniques improved significantly.
Especially with the development of high precision silicon vertex detectors,
the study of $B$~hadrons is now an established part of the physics program
at hadron colliders.

The CDF and D\O~experiments can look back to an already successful
$B$~physics program during the 1992-1996 Run\,I data taking period (for a
review of $B$~physics results from, for example, CDF in Run\,I see
Ref.~\cite{myrevart}).  Nowadays, $B$~physics results from a hadron
collider are fully competitive with the $e^+ e^-$ $B$~factories.  As
discussed later in this review, with the operation of a hadronic track
trigger, CDF reconstructs fully hadronic $B$~decay modes without leptons in
the final state. In many cases, the measurements performed at the Tevatron
Collider are complementary to the $B$~factories. For example, no
$\Bs$~mesons or baryons containing $b$~quarks are produced on the
$\Upsilon(4S)$~resonance.

$B$~hadrons not produced at the $B$~factories are the topics of this
review. We discuss the spectroscopy of excited $B$~states ($B^{**}$,
$B_s^{**}$) and the observation of the \Sb~baryon at the Tevatron.  The
second part of this review discusses the decays of $B$~hadrons and
measurements of branching fractions. We focus on charmless two-body decays
of \Bhh. We end this article by summarizing our finding in the conclusions.

\section{The Tevatron with the CDF \& D\O~Experiments}

The Fermilab accelerator complex has undergone a major upgrade in
preparation for Tevatron Run\,II.  The centre-of-mass energy has been
increased to 1.96~TeV as compared to 1.8~TeV during Run\,I and the Main
Injector, a new 150~GeV proton storage ring, has replaced the Main Ring as
injector of protons and anti-protons into the Tevatron.  The present bunch
crossing time is 396~ns with a $36\times36$ $p\bar p$ bunch operation.  The
luminous region of the Tevatron beam has an RMS of $\sim\!30$~cm~along the
beamline ($z$-direction) with a transverse beamwidth of about 25-30~$\mu$m.

\begin{figure}[tb]
\begin{center}
\centerline{
\epsfig{file=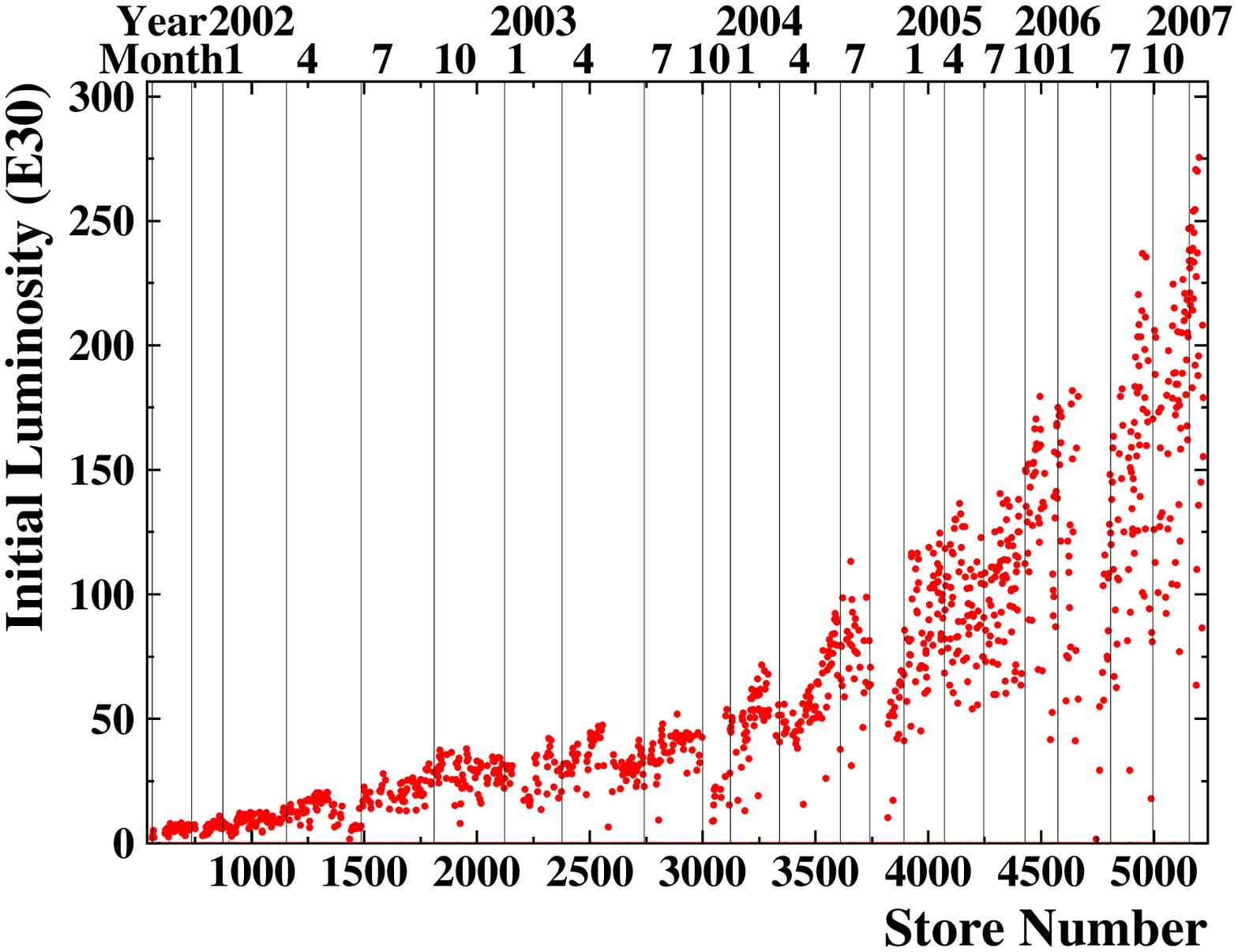,width=3.4in}
\epsfig{file=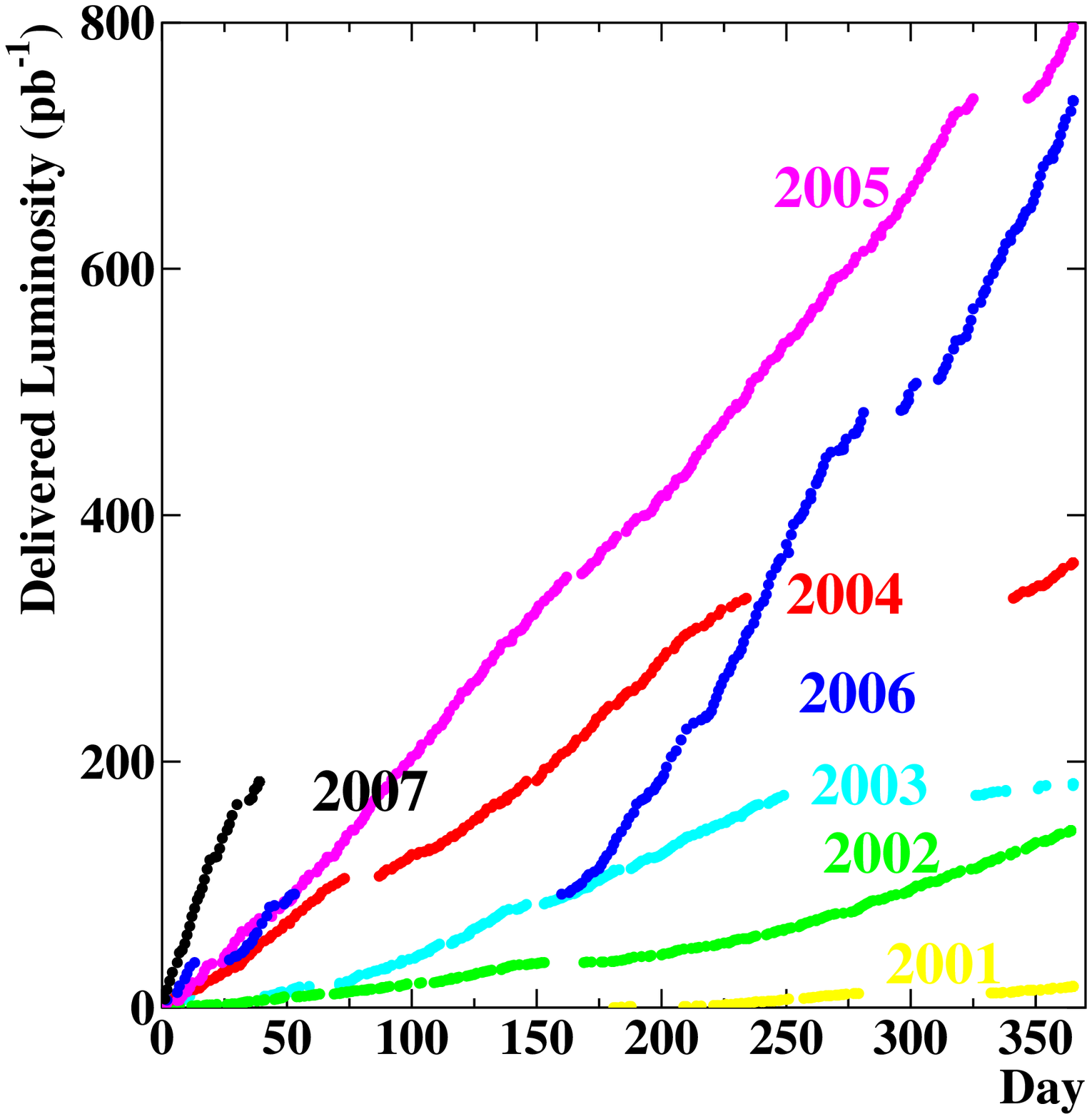,width=2.6in}
\put(-394,150){\large\bf (a)}
\put(-150,160){\large\bf (b)}
}
\caption{Tevatron (a) initial store luminosity from 2002-2006 and (b)
  delivered luminosity per calendar year.} 
\label{fig:TeV}
\end{center}
\end{figure}

The initial Tevatron luminosity steadily increased from 2002 to 2006 as
shown in Figure~\ref{fig:TeV}(a).  By the end of 2006, the peak luminosity
reached by the Tevatron is $>25\cdot 10^{31}$~cm$^{-2}$s$^{-1}$.  The
increase in accelerator performance throughout Run\,II can also be seen by
the delivered luminosity per calendar year as displayed in
Figure~\ref{fig:TeV}(b).  The total integrated luminosity delivered by the
Tevatron to CDF and D\O\ at the end of 2006 is $\sim\!2.2$~fb$^{-1}$ with
about $1.8$~fb$^{-1}$ recorded to tape by each the CDF and D\O~experiments.
However, most results presented in this review use about 1~fb$^{-1}$ of
data.

The CDF detector improvements for Run\,II~\cite{cdfup} were motivated by
the shorter accelerator bunch spacing and the increase in luminosity by an
order of magnitude. All front-end and trigger electronics has been
significantly redesigned and replaced. A DAQ upgrade allows the operation
of a pipelined trigger system. CDF's tracking system was completely renewed
for Run\,II. It consists of a Central Outer Tracker~(COT) with 30\,200
sense wires arranged in 96 layers, between 40 and 137 cm in radius,
organized into eight alternating axial and $\pm 2^{\circ}$ stereo
super-layers.  The transverse momentum resolution is $\sigma_{p_{T}}/p_{T}
\simeq 0.15\%\, p_{T}$/(GeV/$c$).  The specific energy loss by ionization
(\dedx) of charged particles in the COT is measured from the amount of
charge collected by each wire.  The Run\,II silicon vertex detector
consists of seven double sided layers and one single sided layer mounted on
the beam pipe covering a total radial area from 1.5-28~cm. The silicon
vertex detector covers the full Tevatron luminous region and allows for
standalone silicon tracking up to a pseudo-rapidity $|\eta|$ of 2. The
forward calorimeters have been replaced by a new scintillator tile based
plug calorimeter which gives good electron identification up to $|\eta|=2$.
The upgrades to the muon system almost double the central muon coverage and
extend it up to $|\eta|\sim1.5$.  The most important improvements for
$B$~physics in Run\,II are a Silicon Vertex Trigger (SVT) and a
Time-of-Flight (ToF) system with a resolution of about 100~ps. The later
employs 216 three-meter-long scintillator bars located between the outer
radius of the COT and the superconducting solenoid.  The Time-of-Flight
system is most beneficiary for the identification of kaons with a
2\,$\sigma$-separation between $\pi$ and $K$ for $p<1.6$~\gevc.

The D\O~detector also went through a major upgrade before the beginning of
Run\,II~\cite{dup}. The inner tracking system was completely replaced and
includes a new Silicon tracker surrounded by a Scintillating Fiber tracker,
both of which are enclosed in a 2~Tesla solenoidal magnetic
field. Pre-shower counters are located before the uranium/liquid-argon
calorimeter to improve the electron and photon identification. The already
excellent muon system has been further improved by adding more shielding to
reduce beam background.  The Run\,II D\O\ detector has excellent tracking
and lepton acceptance. Tracks with pseudo-rapidity as large as 2.5-3.0
($\theta \approx 10^{\circ}$) and transverse momentum $p_T$ as low as
180~\mevc\ can be reconstructed.  The muon system can identify muons within
$|\eta| < 2.0$. The minimum $p_T$ of the reconstructed muons varies as a
function of $\eta$. In most of the results presented, muons were required
to have $p_T > 2$~\gevc.

\subsection{Triggers for \boldmath{$B$}~Physics}

The total inelastic $p\bar p$ cross section at the Tevatron is about three
orders of magnitude larger than the $b$~quark production cross section. The
CDF and D\O\ trigger system is therefore the most important tool for
finding $B$~decay products. In addition, the cross section for $b$~quark
production is steeply falling. It drops by almost two orders of magnitude
between a $b$~quark $p_T$ of about 8~\gevc\ and 25~\gevc.  To find
$B$~decay products in hadronic collisions, it is desirable to go as low as
possible in the decay products transverse momentum, exploiting as much as
possible of the steeply falling $b$~cross section. Of course, the limiting
factor is the bandwidth of the experiment's data acquisition system.

In Run\,I, all $B$~physics triggers at CDF and D\O\ were based on leptons
including single and dilepton triggers.  In Run\,II, both experiments still
exploit heavy flavour decays which have leptons in the final state.
Identification of dimuon events down to very low momentum is possible,
allowing for efficient $J/\psi \rightarrow \mu^+\mu^-$ triggers. As a
consequence, both experiments are able to fully reconstruct $B$~decay modes
involving $J/\psi$'s.  Both experiments also use inclusive lepton triggers
designed to accept semileptonic $B\rightarrow \ell \nu_\ell X$ decays.
D\O\ has an inclusive muon trigger with excellent acceptance, allowing them
to accumulate very large samples of semileptonic decays.  The CDF
semileptonic triggers require an additional displaced track associated with
the lepton, providing cleaner samples with smaller yields.

In addition, the CDF detector has the ability to select events based upon
track impact parameter.  The Silicon Vertex Trigger gives CDF access to
purely hadronic $B$~decays and makes CDF's $B$~physics program fully
competitive with the one at the $e^+e^-$~$B$~factories. The hadronic track
trigger is the first of its kind operating successfully at a hadron
collider. It works as follows: With a fast track trigger at Level\,1, CDF
finds track pairs in the COT with $p_T>1.5$~\gevc. At Level\,2, these
tracks are linked into the silicon vertex detector and cuts on the track
impact parameter (e.g.~$d > 100$ $\mu$m) are applied.  The SVT track impact
parameter resolution is about 50~$\mu$m including a 33~$\mu$m contribution
from the transverse beam spreading.  The original motivation for CDF's
hadronic track trigger was to select $B^0 \ra \pi\pi$ decays to be used for
$CP$~violation studies.  With the different $B$~trigger strategies above,
the Collider experiments are able to trigger and reconstruct large samples
of heavy flavour hadrons.

\section{Spectroscopy}

\subsection{Study of Orbitally Excited \boldmath{$B$}~Mesons}

The spectroscopy of excited meson states containing $b$~quarks is not well
studied. Only the stable $0^-$ ground states $B^+$, $B^0$ and \Bs\ and the
excited $1^-$ state $B^*$ are established~\cite{PDG}. Quark models predict
the existence of two wide ($B_0^*$ and $B_1^*$) and two narrow ($B_1^0$ and
$B_2^{0*}$) bound $P$-states~\cite{eichten}. The wide states decay through
an $S$-wave and therefore have a large width of a couple of hundred \mevcc,
which makes it difficult to distinguish such states from combinatoric
background. The narrow states decay through a $D$-wave ($L=2$) and thus
should have a small width of around 1~\mevcc~\cite{Ebert,Isguretal}.
Almost all previous observations~\cite{BdsLEP_OPAL,BdsLEP} of the narrow
$P$-states $B_1$ and $B_2^{0*}$ have been made indirectly using inclusive
or semi-exclusive $B$~decays which prevented the separation of both states
and a precise measurement of their properties. In contrast, the masses,
widths and decay branching fractions of these states are predicted with
good precision by the theoretical models~\cite{Ebert,Isguretal}.

$B_1^0$ and $B_2^{0*}$ candidates are reconstructed in the following decay
modes: $B_1^0 \ra B^{*+}\pi^-$ with $B^{*+}\ra B^+\gamma$ and $B_2^{0*} \ra
B^{*+}\pi^-$ with $B^{*+}\ra B^+\gamma$ as well as $B_2^{0*} \ra
B^{+}\pi^-$. In both cases the soft photon from the $B^*$ decay is not
reconstructed resulting in a shift of about 46~\mevcc\ in the mass
spectrum.  D\O\ reconstructs the $B^+$ candidates in the fully
reconstructed mode $B^+\ra J/\psi K^+$ with $J/\psi\ra\mu^+\mu^-$ while CDF
selects $B^+$ mesons in addition through the $B^+\ra D^0\pi^+$ mode with
$D^0\ra K^-\pi^+$. The CDF analysis is based on 360~pb$^{-1}$ of data
resulting in a $B^+\ra J/\psi K^+$ signal of $1867\pm64$ events and
$2182\pm54$ candidates in the $B^+\ra D^0\pi^+$ channel. The D\O\
measurement employs 1~fb$^{-1}$ of Run\,II data and finds a signal peak of
$16\,219\pm180$ events attributed to the decay $B^+\ra J/\psi K^+$.

\begin{figure}[tb]
\begin{center}
\centerline{
\epsfig{file=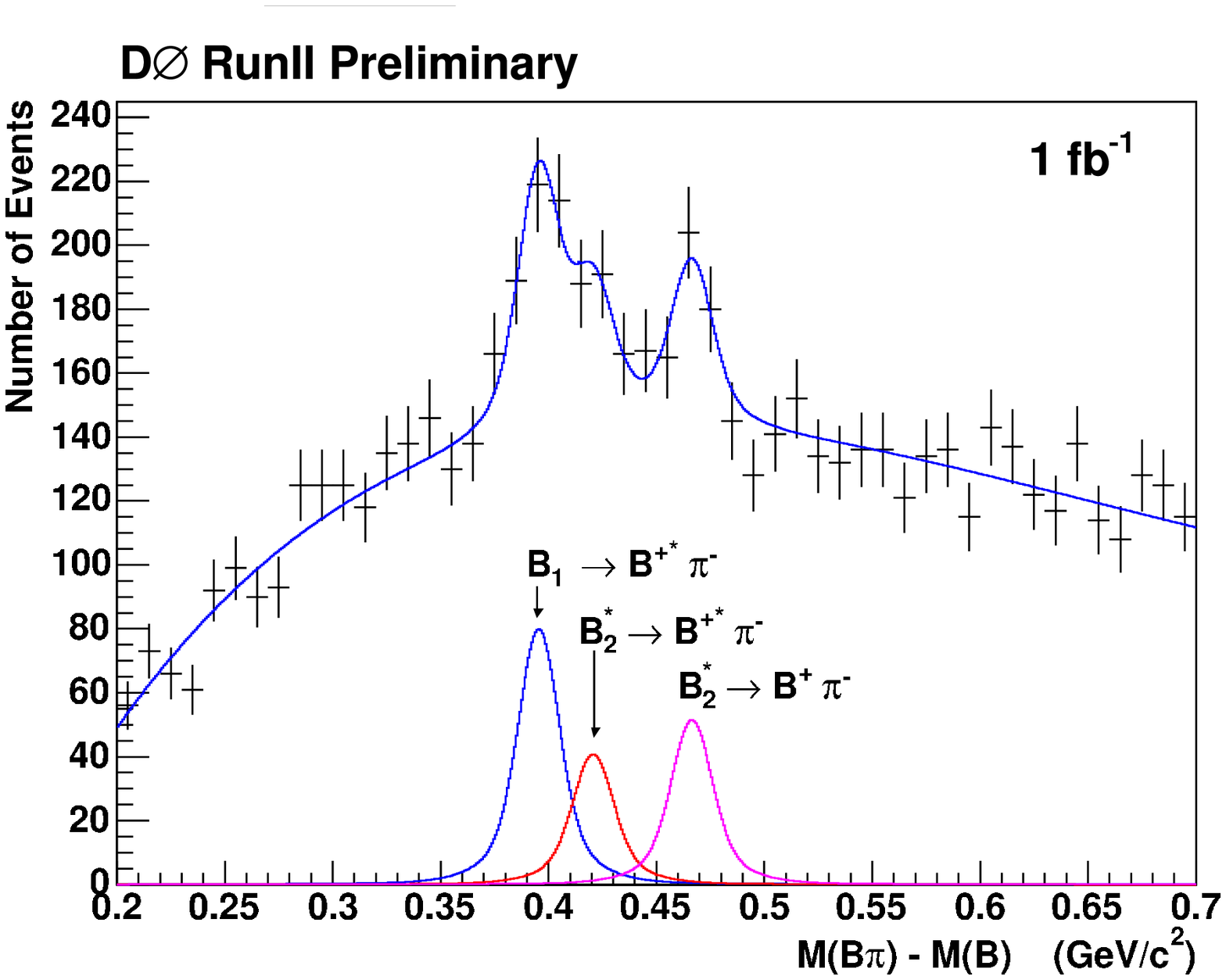,width=3.5in}
\epsfig{file=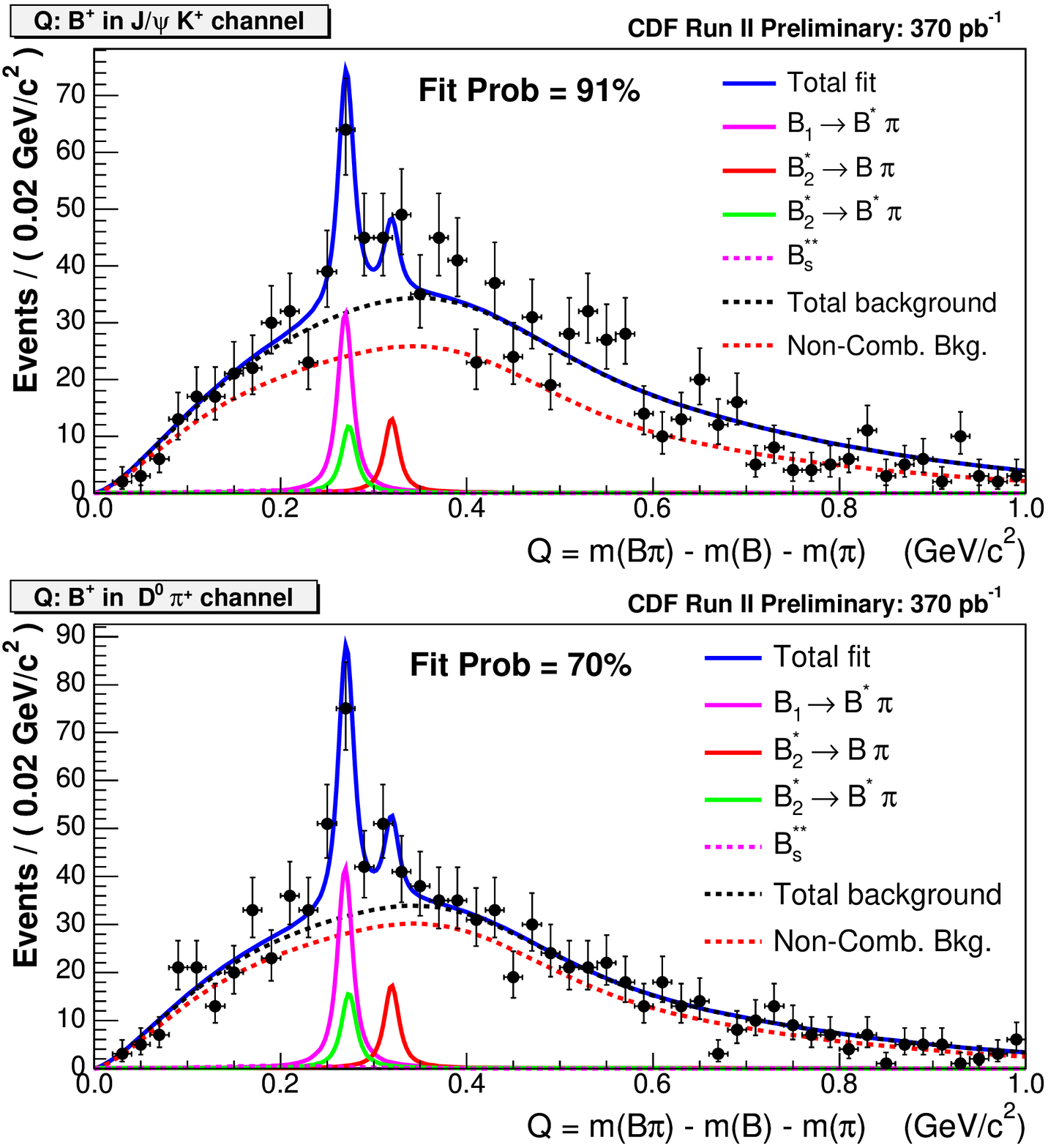,width=2.5in}
\put(-395,150){\large\bf (a)}
\put(-155,170){\large\bf (b)}
\put(-155,70){\large\bf (c)}
}
\caption{Result of the fit to the $B^{**}$ mass difference (a) $\Delta m =
        m(B\pi)-m(B)$ from D\O\ and (b) $Q = m(B\pi)-m(B)-m(\pi)$ from CDF
        in the $B^+\ra J/\psi K^+$ channel and (c) in the $B^+\ra D^0\pi^+$
        mode.}
\label{fig:Bd_double_star}
\end{center}
\end{figure}

D\O\ presents their measured mass distribution as $\Delta m = m(B\pi)-m(B)$
as shown in Figure~\ref{fig:Bd_double_star}(a), while CDF plots $Q =
m(B\pi)-m(B)-m(\pi)$ as displayed in Fig.~\ref{fig:Bd_double_star}(b) and
(c). Clear signals for the narrow excited $B$~states are observed: CDF
reconstructs $80\pm18$ events in $B^+\ra J/\psi K^+$ and $106\pm20$ events
in the $B^+\ra D^0\pi^+$ channel while D\O\ observes a total of $504\pm80$
candidates for the narrow $B^{**}$ states. The measured masses are reported
as $m(B_1^0)=5720.8\pm2.5\pm5.3$~\mevcc\ and
$m(B_2^{0*})-m(B_1^0)=25.2\pm3.0\pm1.1$~\mevcc\ from D\O, while CDF quotes
$m(B_1^0)=5734\pm3\pm2$~\mevcc\ and $m(B_2^{0*})=5738\pm5\pm1$~\mevcc.
Clearly these preliminary results are not in good agreement. CDF currently
works on an update of their analysis using 1~fb$^{-1}$ of data.

\subsection{Observation of Orbitally Excited \boldmath{$B_{sJ}$}~Mesons}

The properties of $\langle b\bar s\rangle$ excited meson states and the
comparison with properties of excited states in the $\langle b\bar
u\rangle$ and $\langle b\bar d\rangle$ systems provide good tests of
various models of quark bound states. These
models~\cite{eichten,Ebert,Falk:1995th} predict the existence of two wide
resonances ($B_{s0}^*$ and $B_{s1}^*$) and two narrow ($B_{s1}^0$ and
$B_{s2}^{0*}$) bound $P$-states. The wide states decay through an $S$-wave
and therefore have a large width of a couple of hundred \mevcc. This makes
it difficult to distinguish such states from combinatoric background. The
narrow states decay through a $D$-wave ($L=2$) and therefore should have a
small width of around 10~\mevcc~\cite{Isguretal}. If the mass of the
$B_{sJ}\ (J=1,2)$ is large enough, then the main decay channel should be
$B^{(*)}K$ as the $\Bs\pi$ decay mode is not allowed by isospin
conservation. Previous observations~\cite{BdsLEP_OPAL} of the narrow
$B_{sJ}$ $P$-states have been made indirectly preventing the separation of
both states.

\begin{figure}[tb]
\begin{center}
\centerline{
\epsfig{file=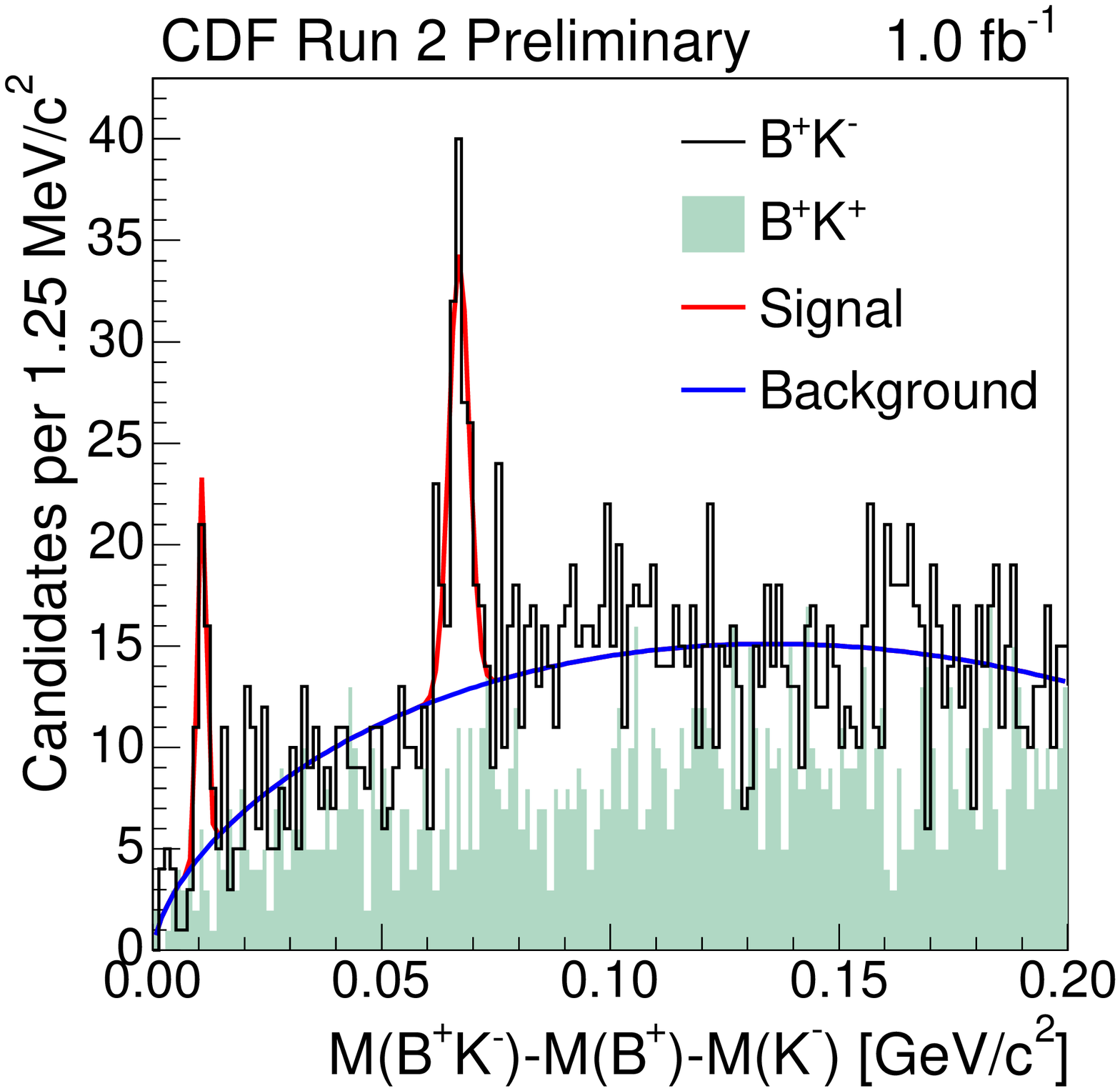,width=2.5in}
\epsfig{file=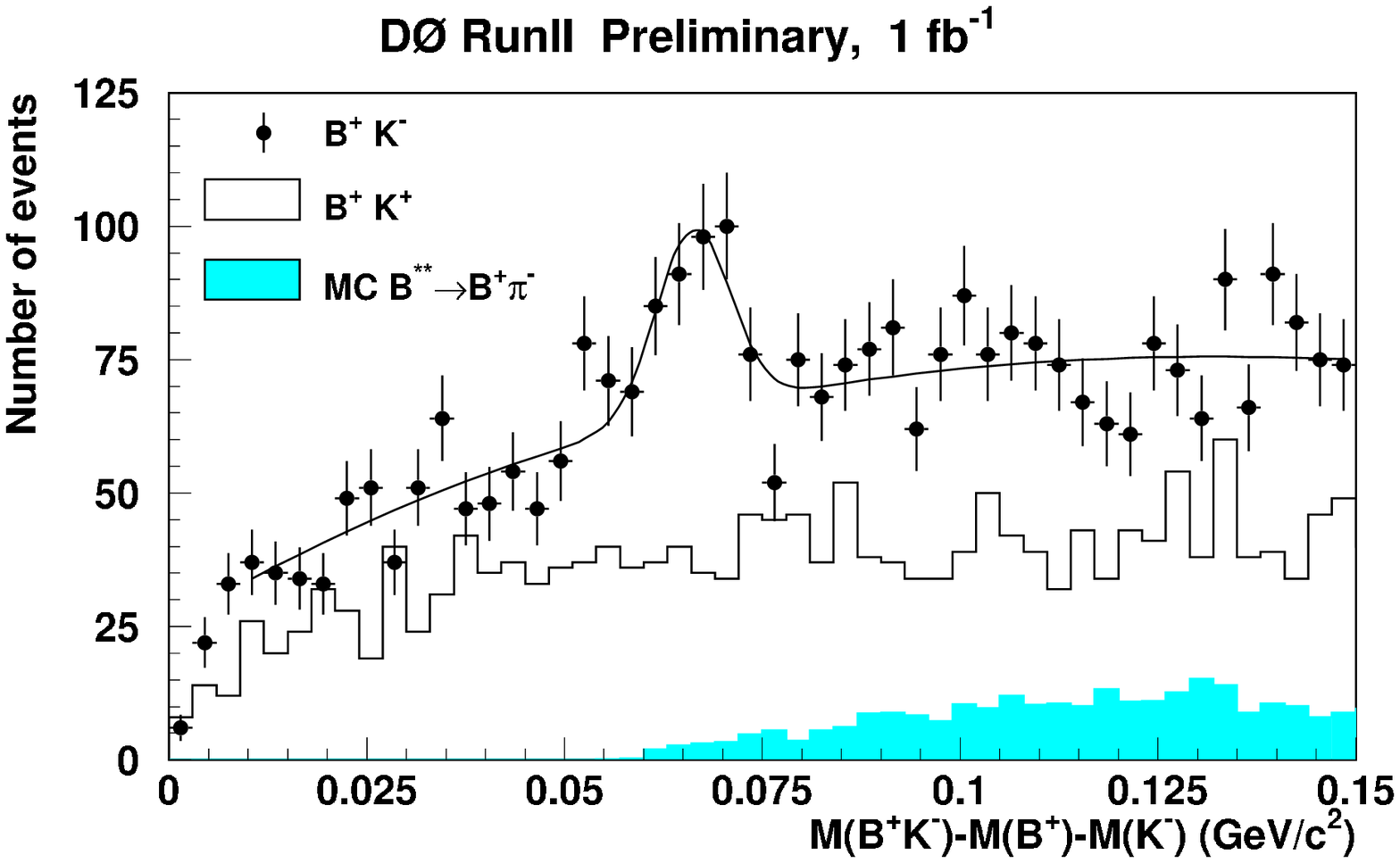,width=3.5in}
\put(-395,140){\large\bf (a)}
\put(-60,120){\large\bf (b)}
}
\caption{Result of the fit to the $B_{sJ}^{**}$ mass difference $Q =
        m(BK)-m(B)-m(K)$ from (a) CDF and (b) D\O.}
\label{fig:Bs_double_star}
\end{center}
\end{figure}

$B_{s1}^0$ and $B_{s2}^{0*}$ candidates are reconstructed in the following
decay modes: $B_{s1}^0 \ra B^{*+}K^-$ with $B^{*+}\ra B^+\gamma$ and
$B_{s2}^{0*} \ra B^{*+}K^-$ with $B^{*+}\ra B^+\gamma$ as well as
$B_{s2}^{0*} \ra B^{+}K^-$. In both cases the soft photon from the $B^*$
decay is not reconstructed resulting in a shift in the mass spectrum.  D\O\
reconstructs the $B^+$ candidates in the fully reconstructed mode $B^+\ra
J/\psi K^+$ with $J/\psi\ra\mu^+\mu^-$ while CDF selects $B^+$ mesons in
addition through the $B^+\ra D^0\pi^+$ mode with $D^0\ra K^-\pi^+$. The CDF
and D\O\ measurements are each based on 1~fb$^{-1}$ of Run\,II data.  The
CDF analysis finds $\sim31\,000$~$B^+\ra J/\psi K^+$ events and
$\sim27\,200$ candidates in the $B^+\ra D^0\pi^+$ channel.  D\O\ uses a
signal of $16\,219\pm180$ $B^+$ events from the decay $B^+\ra J/\psi K^+$.
Both experiments present their measured mass distribution in the quantity
$Q = m(BK)-m(B)-m(K)$ as displayed in Figure~\ref{fig:Bs_double_star}(a)
and (b).

A clear signal at $Q\sim 67$~\mevcc\ is observed by CDF and D\O\ (see
Fig.~\ref{fig:Bs_double_star}), which is interpreted as the $B_{s2}^{0*}$
state.  CDF reconstructs $95\pm23$ events in the peak at $Q=67.0~\mevcc$
while D\O\ reports $135\pm31$ events at $Q=66.4\pm1.4~\mevcc$.  In
addition, CDF observes $36\pm9$ events in a peak at $Q\sim10.7~\mevcc$
which is interpreted as first evidence for the $B_{s1}^0$~state.  The
measured masses are reported as $m(B_{s2}^{0*})=5839.1\pm1.4\pm1.5$~\mevcc\
from D\O, while CDF quotes $m(B_{s1}^0)=5829.4\pm0.2\pm0.6$~\mevcc\ and
$m(B_{s2}^{0*})=5839.6\pm0.4\pm0.5$~\mevcc. The results from CDF and D\O\
are in good agreement.

\subsection{Observation of \boldmath{$\Sib$} Baryons}

Until recently only one bottom baryon, the \Lb, has been directly observed.
At present the CDF collaboration has accumulated the world's largest data
sample of bottom baryons, due to a combination of two factors -- the CDF
displaced track trigger, and the $\sim {\rm 1~fb^{-1}}$ of integrated
luminosity delivered by the Tevatron.  Using a sample of fully
reconstructed $\Lb\ra\Lc\pi^-$ candidates collected with the displaced
track trigger, CDF searched for the decay $\Sib^{(*)\pm} \ra \Lb\pi^{\pm}$.

The QCD treatment of quark-quark interactions significantly simplifies if
one of the participating quarks is much heavier than the QCD confinement
scale $\Lambda_{\rm QCD}$.  In the limit of ${m_Q\ra\infty}$, where ${m_Q}$
is the mass of the heavy quark, the angular momentum and flavour of the
light quark become good quantum numbers. This approach, known as Heavy
Quark Effective Theory (HQET), thus views a baryon made out of one heavy
quark and two light quarks as consisting of a heavy static color field
surrounded by a cloud corresponding to the light diquark system.  In SU(3)
the two quarks are in diquark form $\bar{3}$ and $6$ according to the
decomposition $3 \otimes 3 = \bar{3} \oplus 6$, leading to a generic scheme
of baryon classification. Diquark states containing quarks in an
antisymmetric flavour configuration, $[q_1,q_2]$, are called $\Lambda$-type
whereas states with diquarks containing quarks in a flavour symmetric
state, $\{q_1,q_2\}$, are called $\Sigma$-type.

In the $\Sigma$-type ground state the light diquark system has isospin
$I=1$ and $J^P=1^+$.  Together with the heavy quark this leads to a doublet
of baryons with $J^P=\frac{1}{2}^+$ (\Sib) and $J^P=\frac{3}{2}^+$
($\Sib^*$).  The ground state $\Sigma$-type baryons decay strongly to
$\Lambda$-type baryons by emitting pions.  In the limit ${m_Q\ra\infty}$,
the spin doublet $\{\Sib,\Sib^*\}$ would be exactly degenerate since an
infinitely heavy quark does not have a spin interaction with a light
diquark system.  As the heavy quark is not infinitely massive, there will
be a small mass splitting between the doublet states and there is an
additional isospin splitting between the $\Sib^{(*)-}$ and $\Sib^{(*)+}$
states~\cite{Rosner:2006yk}.  There exist a number of predictions for the
masses and isospin splittings of these states using HQET, non-relativistic
and relativistic potential models, $1/{\rm N}_c$ expansion, sum rules and
lattice QCD. References~\cite{Rosner:2006yk,Stanley:1980fe} contain some of
the existing theoretical estimates, while Table~\ref{tab:theory} summarizes
the range of predictions. The natural width of \Sib~baryons is expected to
be dominated by single pion transitions.  Decays of the type
$\Sigma_{c,b}\ra\Lambda_{c,b}\gamma$ are expected to have significantly
smaller ($\sim 100~{\rm keV}/c^2$) partial widths than the single pion
transition, and are thus negligible.  The partial width of the $P$-wave
one-pion transition thus depends on the available phase space.

\begin{table}[tb]
\begin{center}
\begin{tabular}{|l|c|} \hline
\Sib~property & Expected value [$\mevcc$] \\ 
\hline m(\Sib) - m(\Lb) & 180 - 210 \\ 
$m(\Sib^*) - m(\Sib)$ & 10 - 40 \\ 
$m(\Sib^-) - m(\Sib^+)$ & 5 - 7 \\ 
$\Gamma(\Sib)$, $\Gamma(\Sib^*)$ & $\sim$8, $\sim$15 \\ 
\hline
\end{tabular}
\caption{General range of theoretical predictions for the $\Sib^{(*)\pm}$
  states from References~\cite{Rosner:2006yk,Stanley:1980fe}.
\label{tab:theory}}
\end{center}
\end{table}

In analogy with the $B$~meson hadronization chain, in this analysis events
are separated into ``same charge'' or SC and ``opposite charge'' or OC
combinations.  As the \Lb\ is neutral, the charge of the soft pion track
determines the charge of the \Sib~baryon, and there will be \Sib~signals
for both positive and negative pions.  SC (OC) is defined as events where
the \Sib~pion has the same (opposite) charge as the pion from the
\Lb~decay.  With these definitions, the SC distribution contains all
$\Sigma_b^{(*)-}$ and $\bar{\Sigma}_b^{(*)-}$ candidates while OC contains
$\Sigma_b^{(*)+}$ and $\bar{\Sigma}_b^{(*)+}$.

\begin{figure}[tb]
\begin{center}
\centerline{
\epsfig{file=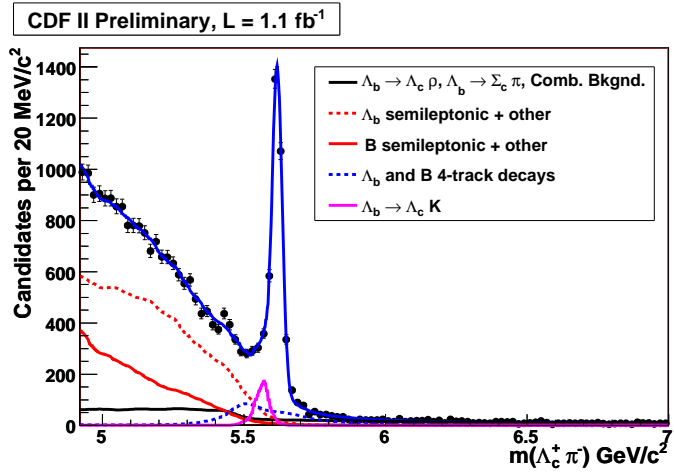,width=3.5in}
}
\caption{Fit to the invariant mass of $\Lb\ra\Lc\pi^-$ candidates. The
solid blue line is the total fit, while the primary background sources are
listed in the legend.}
\label{fig:SigmaB_LambdaB}
\end{center}
\end{figure}

The present analysis is based on events collected by the CDF detector from
2002 through February 2006, with an integrated luminosity of
\(\mathcal{L}=1070 \pm 60~\mbox{pb}^{-1}\).  Events collected on the two
track trigger are used to reconstruct the decay chain $\Lb \ra \Lc \pi,\
\Lc \ra p K^- \pi^+$.  CDF reconstructs a \Lb~yield of approximately 2800
candidates in the signal region $m(\Lb)\in [5.565, 5.670]~\gevcc$, with the
\Lb~mass plot shown in Figure~\ref{fig:SigmaB_LambdaB}.
 
To separate out the resolution on the mass of each \Lb~candidate, CDF
searches for narrow resonances in the mass difference distribution of $Q =
m(\Lb \pi) - m(\Lb) - m_\pi$.  Unless explicitly stated, \Sib~refers to
both the $J=\frac{1}{2}$ ($\Sib^{\pm}$) and $J=\frac{3}{2}$ ($\Sib^{*\pm}$)
states.  There is no transverse momentum cut applied to the pion from the
\Sib~decay, since these tracks are expected to be very soft.  In order to
perform an unbiased search, the cuts for the \Sib~reconstruction are
optimized first with the \Sib~signal region blinded.  From theoretical
predictions the \Sib~signal region is chosen as $30 < Q < 100~\mevcc$,
while the upper and lower sideband regions of $0 < Q < 30~\mevcc$ and $100
< Q < 500~\mevcc$ represent the \Sib~background.  The signal for the
optimization is taken from a PYTHIA Monte Carlo \Sib~sample, with the
decays $\Sib\ra\Lb\pi,\ \Lb\ra\Lc\pi^-,\ \Lc\ra pK^-\pi^+$ forced.

The backgrounds under the \Lb~signal region in the \Lb~mass distribution
will also be present in the \Sib~$Q$-distribution.  The primary sources of
background are \Lb~hadronization and underlying event, hadronization and
underlying event of other $B$~meson reflections and combinatorial
background underneath the \Lb~peak. The percentage of each background
component in the \Lb~signal region is derived from the \Lb~mass fit, and is
determined as 86\%~\Lb~signal, 9\%~backgrounds and 5\% combinatorial
background.  Other backgrounds ({\it e.g.} from 5-track decays where one
track is taken as the $\pi_{\Sib}$ candidate) are negligible, as confirmed
in inclusive single-$b$-hadron Monte Carlo samples.

\begin{figure}[tb]
\begin{center}
\centerline{
\epsfig{file=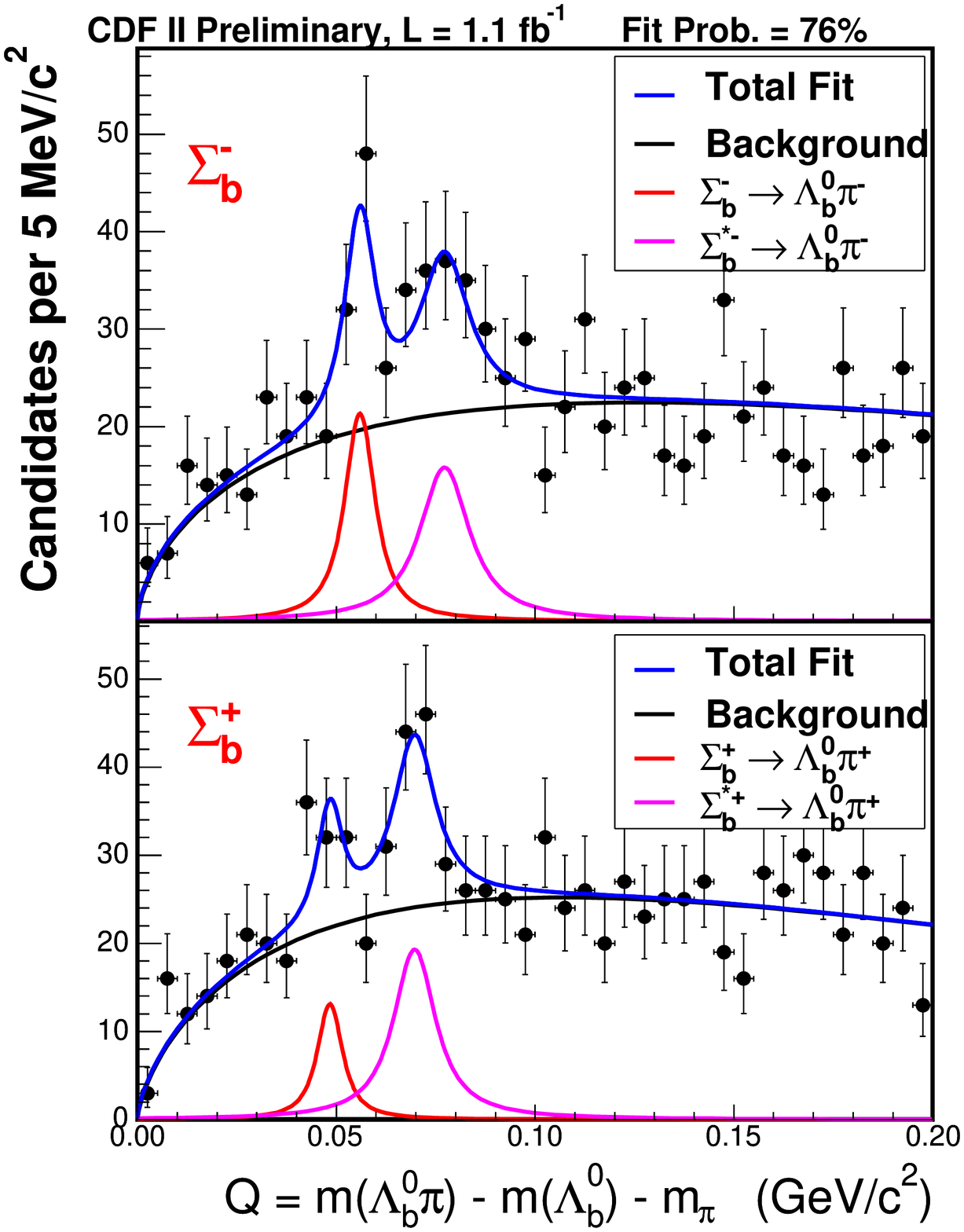,width=3.0in}
\epsfig{file=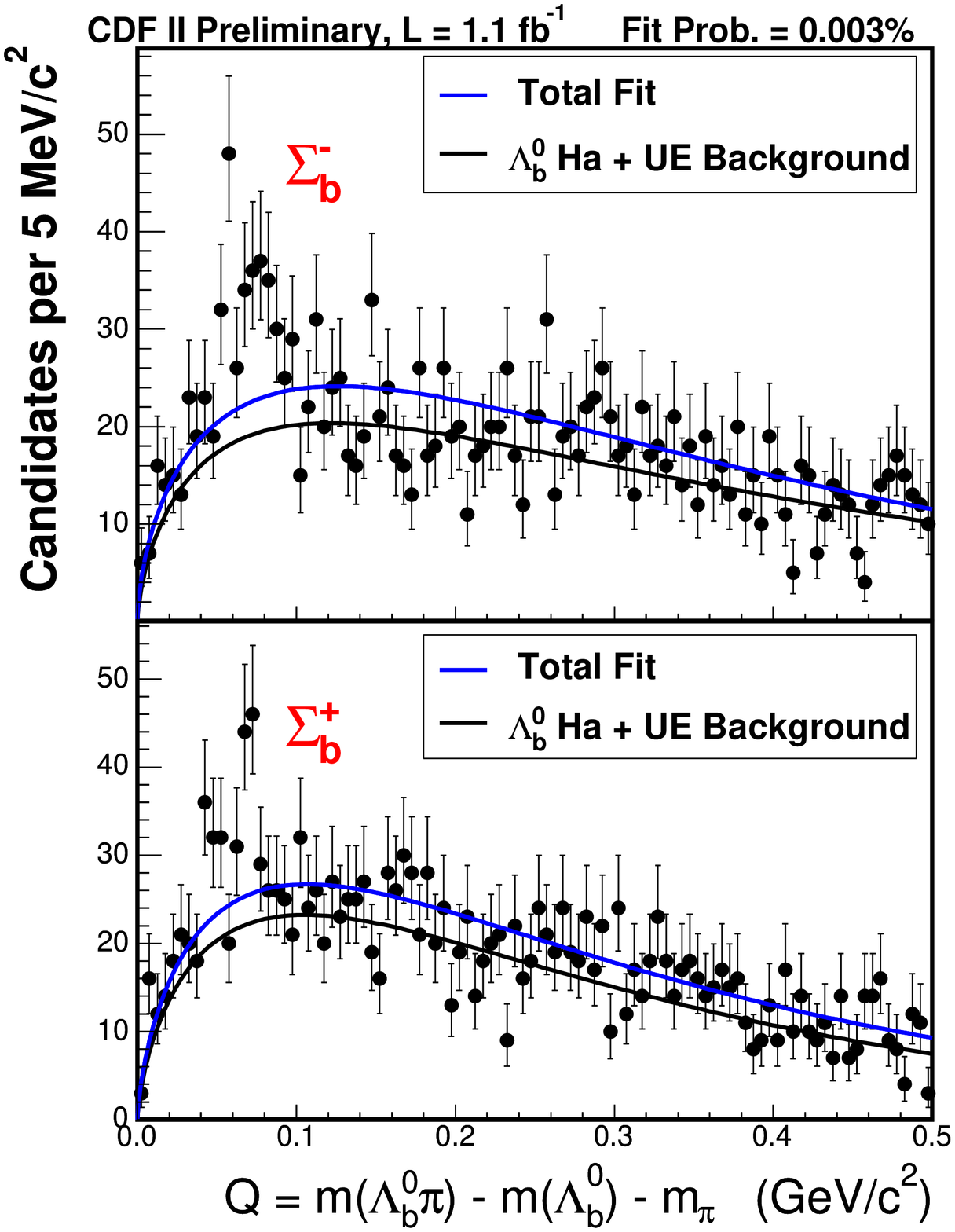,width=3.0in}
\put(-330,240){\large\bf (a)}
\put(-50,205){\large\bf (b)}
}
\caption{(a) Simultaneous fit to the \Sib~states and (b) with an alternate
  signal description assuming no signal is present (null hypothesis). }
\label{fig:SigmaB_result}
\end{center}
\end{figure}

Upon unblinding the $Q$ signal region, there is an excess observed in data
over predicted backgrounds.  The excess over background is shown in
Table~\ref{tab:bin_count}.  CDF performs a simultaneous unbinned likelihood
fit to SC~and OC~data.  To the already described background components,
four peaks are added, one for each of the expected \Sib~states.  Each peak
is a sum of two Breit-Wigner shapes, each convoluted with two Gaussian
resolution functions.  The detector resolution has a dominant narrow core
and a small broader shape describing the tails where the PDF for each peak
takes both into account.  Due to low statistics, CDF constrains
$m(\Sib^{*+})-m(\Sib^+)$ and $m(\Sib^{*-})-m(\Sib^-)$ to be the same.  The
results of the fit are given in Tab.~\ref{tab:fit_results} and displayed in
Fig.~\ref{fig:SigmaB_result}(a).

\begin{table}[tb]
\begin{center}
\begin{tabular}{|l|c|c|c|} 
\hline
Sample & Data events & Bkg events & Data excess over bkg \\
\hline
Same charge & 416 & 268 & 148 \\
Opposite charge & 406 & 298 & 108 \\
\hline
\end{tabular}
\caption{Summary of the number of events in the $Q$ signal region
        ($Q$ $\in$ [0.03, 0.1]~\gevcc) for data and 
        predicted background.}
\label{tab:bin_count}
\end{center}
\end{table}

\begin{table}
\begin{center}
\begin{tabular}{|l|c|c|c|} 
\hline
Parameter                     & Value         & Parabolic Error & MINOS Errors
 \\
\hline
$Q(\Sib^+)$ ($\mevcc$)  & 48.4          & 2.02     & (+2.02, -2.29) \\
$Q(\Sib^-)$ ($\mevcc$)  & 55.9          & 0.963    & (+0.990, -0.959) \\
$Q(\Sib^*)$ - $Q(\Sib)$ ($\mevcc$) & 21.3          & 1.93     & (+2.03, -1.
94) \\
\hline
$\Sib^+$ events                 & 29            & 12.0     & (+12.4, -11.6) \\
$\Sib^-$ events                 & 60            & 14.3     & (+14.8, -13.8) \\
$\Sib^{*+}$ events              & 74            & 16.8     & (+17.2, -16.3) \\
$\Sib^{*-}$ events              & 74            & 17.8     & (+18.2, -17.4) \\
\hline
-ln(Likelihood)                 & -24553.5      & --       & --             \\
\hline
\end{tabular}
\caption{Fit parameters and error values from the fit to data.  Positive
        and negative errors are quoted separately as the error range is 
        asymmetric.}
\label{tab:fit_results}
\end{center}
\end{table}

All systematic uncertainties on the mass difference measurements are small
compared to their statistical errors.  The systematic errors from the
tracking sources are determined by comparing the mean and the width of the
peak in $m(D^{*+})-m(D^0)$ between data and Monte Carlo simulation split up
in several regions of track $p_T$.  The largest discrepancy of the $D^{*+}$
peak is $0.06~\mevcc$ which is taken as the systematic error for all four
peaks.  The discrepancy in the mass resolution could be as large as
$20\%$. The effect of a broader resolution is evaluated via a sample of Toy
Monte Carlo experiments.  The remaining systematics come from assumptions
made in the fit to the data, such as the use of fixed background shapes.
For the parameters associated with an individual systematic uncertainty,
Toy MC samples are generated where these parameters are varied.  The sample
is then fit with both the default fit and the fit with varied parameters.
The difference between fit parameter values in the varied fit and the
default fit is caused by the systematic variation and constitutes the
associated systematic error.

To evaluate the significance of the measurement, the null hypothesis is
tested. The data is fit with no signal and with the standard fit using four
peaks.  Then the likelihood ratio is computed as $LR = {L_1}/{L_2}$, where
$L_2$ is the four signal peak hypothesis and $L_1$ is the corresponding
hypothesis with no peaks. The result of this fit is shown in
Figure~\ref{fig:SigmaB_result}(b) and a likelihood ratio of $\sim10^{-19}$
is obtained indicating the observation of the $\Sib^{(*)\pm}$~states.

To summarize, the lowest lying charged $\Lb\pi$ resonant states are
observed in 1~fb$^{-1}$ of data collected by the CDF detector.  These are
consistent with the lowest lying charged $\Sib^{(*)\pm}$ baryons.  The $Q$
values of $\Sib^-$ and $\Sib^+$, and the $\Sib^*$-\Sib~mass difference, are
measured to be:
\\ - $m(\Sib^-)-m(\Lb)-m(\pi)=55.9\pm1.0$ (stat)
        $\pm 0.1$ (syst)~\mevcc, 
\\ - $m(\Sib^+)-m(\Lb)-m(\pi)=48.4^{+2.0}_{-2.3}$ (stat)
        $\pm 0.1$ (syst)~\mevcc, 
\\ - $m(\Sib^{*-})-m(\Sib^-)=m(\Sib^{*+})-m(\Sib^+)
        = 21.3^{+2.0}_{-1.9}$ (stat) $^{+0.4}_{-0.2}$ (syst)~\mevcc.

Using the best CDF mass measurement for the \Lb~mass, which is
$m(\Lb)=5619.7\pm1.2$ (stat) $\pm1.2$ (syst)~\mevcc, the absolute mass
values and number of events are:
\\ - $m(\Sib^+)=5808^{+2.0}_{-2.3}$ (stat) $\pm 1.7$ (syst)~\mevcc,
  N($\Sib^+)= 29^{+12.4}_{-11.6}$ (stat) $^{+5.0}_{-3.4}$ (syst),
\\ - $m(\Sib^-)=5816^{+1.0}_{-1.0}$ (stat) $\pm 1.7$ (syst)~\mevcc,
  N($\Sib^-)=60^{+14.8}_{-13.8}$ (stat) $^{+8.4}_{-4.0}$ (syst),
\\ - $m(\Sib^{*+})=5829^{+1.6}_{-1.8}$ (stat) $\pm 1.7$ (syst)~\mevcc,
  N($\Sib^{*+})=74^{+17.2}_{-16.3}$ (stat) $^{+10.3}_{-5.7}$ (syst),
\\ - $m(\Sib^{*-})=5837^{+2.1}_{-1.9}$ (stat) $\pm 1.7$ (syst)~\mevcc,
  N($\Sib^{*-})=74^{+18.2}_{-17.4}$ (stat) $^{+15.6}_{-5.0}$ (syst).

\section{Decay of \boldmath{$B$} Hadrons}

In this Section we focus on a new CDF result involving the branching
fractions and time-integrated direct $CP$~asymmetries for \Bd\ and \Bs\
decay modes into pairs of charmless charged hadrons \Bhh.

\subsection{Results from Charmless Two-Body Decays \boldmath{$\Bhh$}}

The decay modes of $B$~mesons into pairs of charmless pseudo-scalar mesons
are effective probes of the quark-mixing matrix (CKM) and sensitive to
potential new physics effects.  The large production rate of $B$~hadrons at
the Tevatron allows measuring such decays in new modes, which are important
to supplement our understanding of $B$~meson decays.  The still unobserved
\BsKpi~decay mode could be used to measure the angle
$\gamma$~\cite{Gronau:2000md} of the CKM unitarity triangle and its
$CP$~asymmetry could be a powerful model-independent test of the source of
direct $CP$~violation in the $B$~meson system \cite{Lipkin-BsKpi}. This may
provide useful information to solve the current discrepancy between the
asymmetries observed in the neutral and charged $B$~modes~\cite{HFAG06}.
The \Bspipi\ and \BdKK\ decay channels proceed only through annihilation
diagrams, which are currently poorly known and constitute a source of
significant uncertainty in many theoretical calculations~\cite{B-N,Bspipi}.
A measurement of both modes would allow a determination of the strength of
penguin-annihilation diagrams~\cite{Burasetal}.

\subsubsection{Data Selection}

CDF analysed a sample (integrated luminosity ${\cal L}\sim1$~fb$^{-1}$) of
pairs of oppositely charged particles with $p_{T} > 2$~\gevc\ and $p_{T}(1)
+ p_{T}(2) > 5.5$~\gevc, used to form $B^0_{(s)}$~meson candidates. In
addition, the trigger required a transverse opening-angle $20^\circ <
\Delta\phi < 135^\circ$ between the two tracks, to reject background from
particle pairs within the same jet and from back-to-back jets.  In
addition, both charged particles are required to originate from a displaced
vertex with a large impact parameter $d_0$ (100 $\mu$m $< d_0 < 1$~mm),
while the $B^0_{(s)}$~meson candidate is required to be produced in the
primary $\bar{p}p$ interaction ($d_0(B)< 140$~$\mu$m) and to have traveled
a transverse distance $L_{xy}(B)>200$~$\mu$m.

In the offline analysis, an unbiased optimization procedure determines a
tightened selection on track-pairs fit to a common decay-vertex.  CDF
chooses selection cuts minimizing directly the expected uncertainty
(through several pseudo-experiments) of the physics observables to be
measured.  CDF decided to use two different sets of cuts, optimizing
separately the measurements of \acpbdkpi\ and \br(\BsKpi).  For the latter,
the sensitivity for discovery and limit setting~\cite{gp0308063} was
optimized rather than the statistical uncertainty on the particular
observational parameter, since this mode had not yet been observed.  It is
verified that the former set of cuts is also adequate to measure other
decay rates of the larger yield modes (\Bdpipi, \BsKK), while the latter,
tighter set of cuts, is well suited to measure the decay rates and
$CP$~asymmetries related to the rare modes (\Bspipi, \BdKK, \Lbppi, \LbpK).

\begin{figure}[tb]
\begin{center}
\centerline{
\epsfig{file=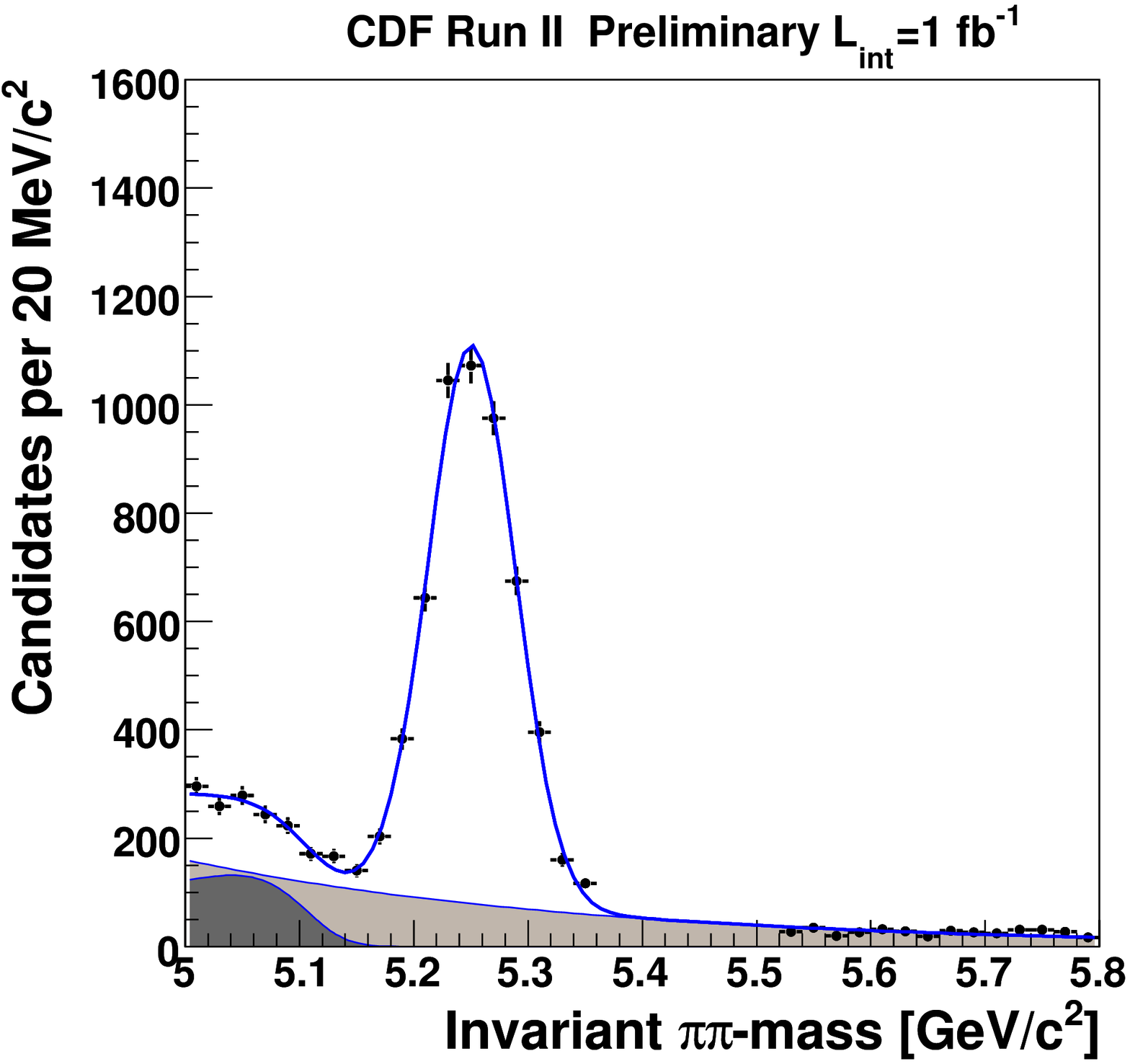,width=3.0in}
\epsfig{file=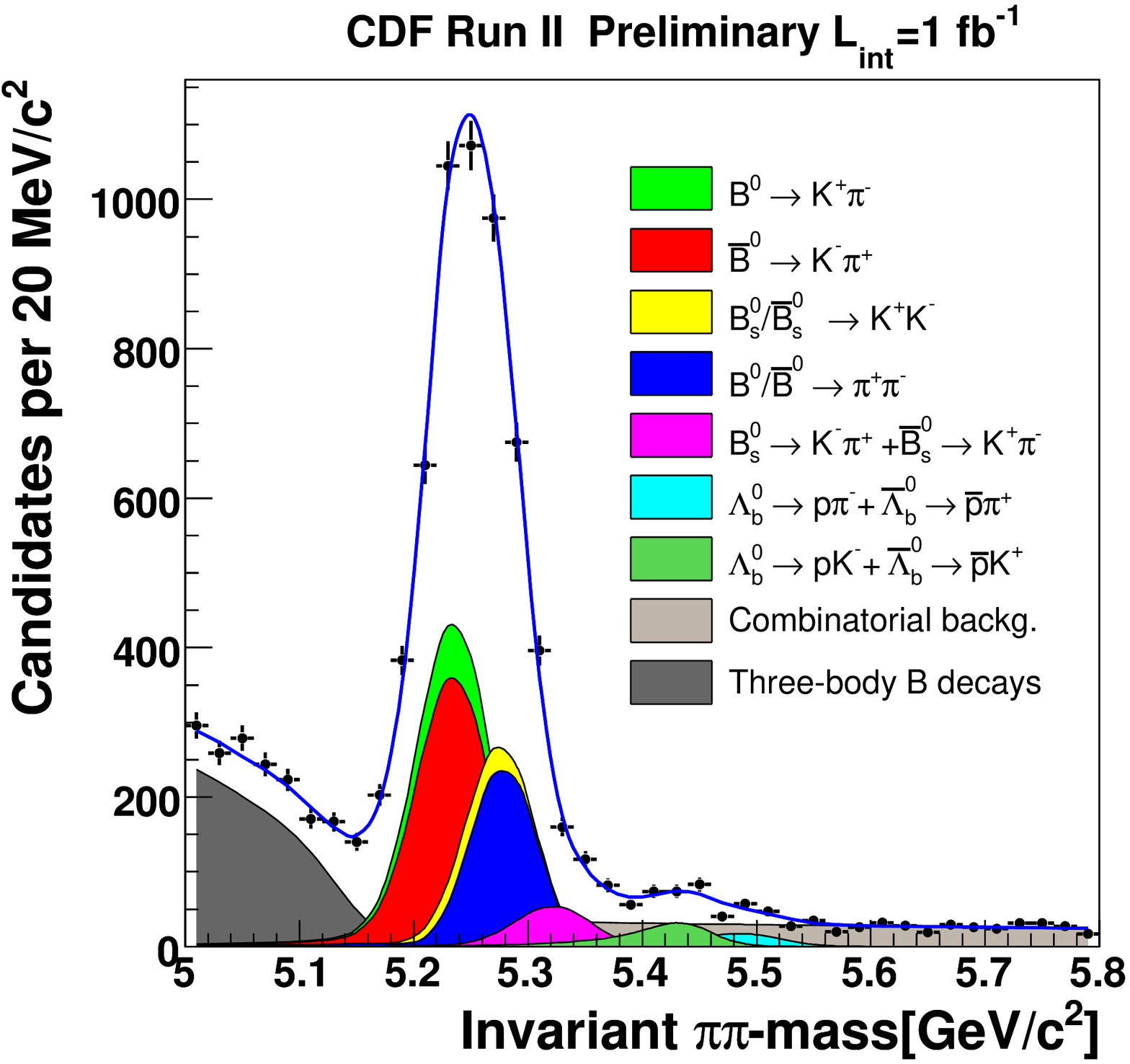,width=3.0in}
\put(-265,170){\large\bf (a)}
\put(-170,170){\large\bf (b)}
}
\caption{(a) Invariant mass distribution of \Bhh\ candidates passing all
selection requirements optimized to measure \br(\BsKpi), using the pion mass
assumption for both decay products. The cumulative projections of the
likelihood fit for each mode are overlaid in (b).}
\label{fig:projections}
\end{center}
\end{figure}

In addition to tightening the trigger cuts in the offline analysis, other
discriminating variables such as the isolation of the $B^0_{(s)}$~meson and
the information provided by the 3D reconstruction capability of the CDF
tracking system are used, allowing a great improvement in the signal
purity.  Isolation is defined as $I(B)= p_T(B)/[p_T(B) + \sum_{i} p_T(i)]$,
in which the sum runs over every other track within a cone of radius one in
the $\eta-\phi$ space around the $B^0_{(s)}$~meson flight-direction. By
requiring $I(B)> 0.5$ the background is reduced by a factor four while
keeping almost 80\% of the $B$~signal. The 3D silicon tracking allows to
resolve multiple vertices along the beam direction and to reject fake
tracks reducing the background by another factor of two, with small
inefficiency on the signal.  The resulting $\pi\pi$~invariant mass
distribution shown in Figure~\ref{fig:projections}(a) display a clean
signal of \Bhh\ decays. In spite of a good mass resolution ($\approx
22~\mevcc$), the various \Bhh\ modes overlap into an unresolved mass peak.

\subsubsection{Fit of Sample Composition}

The resolution in invariant mass and in particle identification is not
sufficient for separating individual decay modes on an event-by-event
basis. Therefore CDF performs an unbinned maximum likelihood fit, combining
kinematic and particle identification information, to statistically
determine the contribution of each mode and the $CP$~asymmetries.  For the
kinematic portion, CDF uses three loosely correlated observables to
summarize the information carried by all possible values of invariant mass
of the $B$~candidate, resulting in different mass assignments to the two
outgoing particles. These are: (a) the mass $m_{\pi\pi}$ calculated with
the charged pion mass assignment to both particles, (b) the signed momentum
imbalance $\alpha = (1-p_1/p_2) q_{1}$, where $p_1$ ($p_2$) is the lower
(higher) of the particle momenta, and $q_1$ is the sign of the charge of
the particle of momentum $p_{1}$, and (c) the scalar sum of the particle
momenta $p_{tot}=p_1 + p_2$.  Using these three variables, the mass of any
particular mode $m_{12}$ can be written as:
\begin{equation}\label{eq:Mpipi2}
m^{2}_{12} = m^{2}_{\pi\pi}  -  2 m_{\pi}^2 +m_{1}^2+m_{2}^2
- 2 \sqrt{p_{1}^2+m_{\pi}^2} \sqrt{p_{2}^2+m_{\pi}^2} 
-  2\sqrt{p_{1}^2+m_{1}^2} \sqrt{p_{2}^2+m_{2}^2},
\end{equation}
\vspace{-0.5cm}
\begin{equation}\label{eq:sostituzione}
p_1 = \frac{1-|\alpha|}{2-|\alpha|}\,p_{tot},\ \ \ \ \ 
p_2 = \frac{1}{2-|\alpha|}\,p_{tot},
\end{equation}
where $m_{1}$ ($m_{2}$) is the mass of the lower (higher) momentum
particle. For simplicity Eq.~(\ref{eq:Mpipi2}) is written as a function
of $p_{1}$ and $p_{2}$ instead of $\alpha$ and $p_{tot}$ but in the
likelihood fit it is used as a function of $\alpha$ and $p_{tot}$.

\begin{figure}[tb]
\begin{center}
\centerline{ \epsfig{file=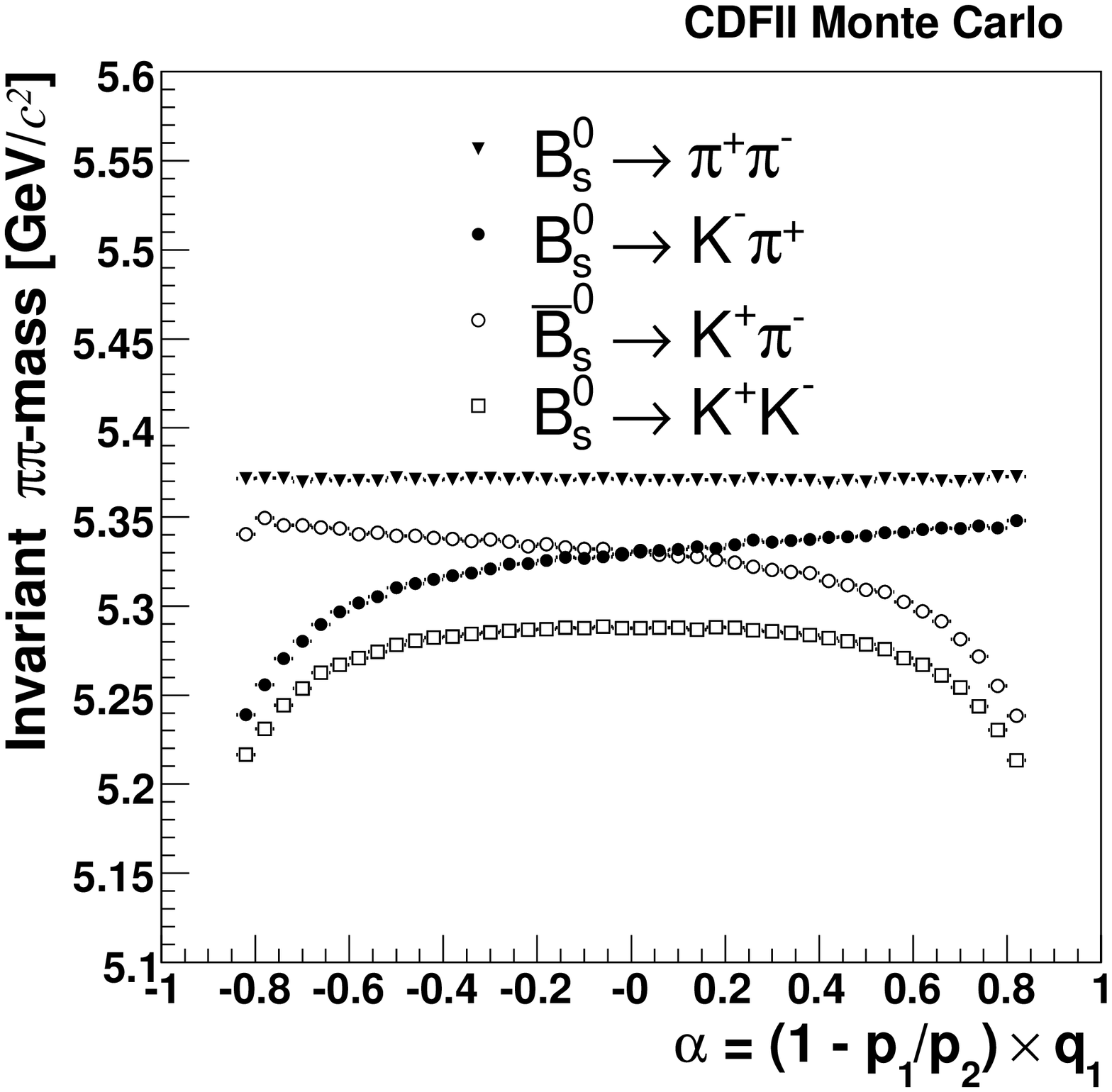,width=3.0in}
\epsfig{file=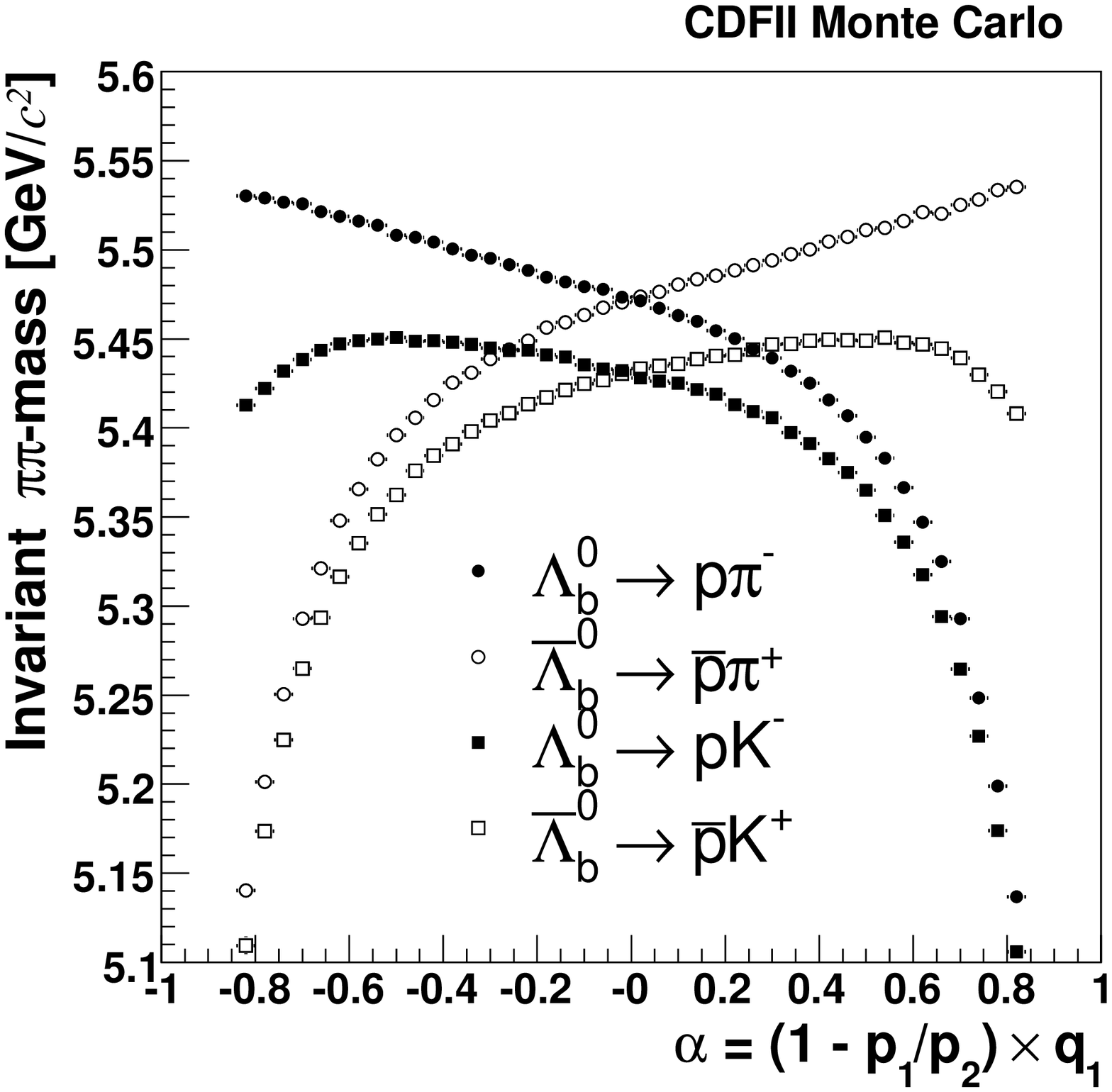,width=3.0in} \put(-265,170){\large\bf (a)}
\put(-120,170){\large\bf (b)} }
\caption{Average $m_{\pi\pi}$ vs $\alpha$ for simulated samples of (a) \Bs\
and (b) \Lb\ candidates, where self-tagging final states ($K^+\pi^-$ and
$K^-\pi^+$, $ph^-$ and $\bar{p}h^+$) are treated separately. The
corresponding plots for the \Bd\ are similar to \Bs\, but shifted for the
mass difference.}
\label{fig:mpipi_vs_alpha}
\end{center}
\end{figure}

\begin{figure}[tb]
\begin{center}
\centerline{
\epsfig{file=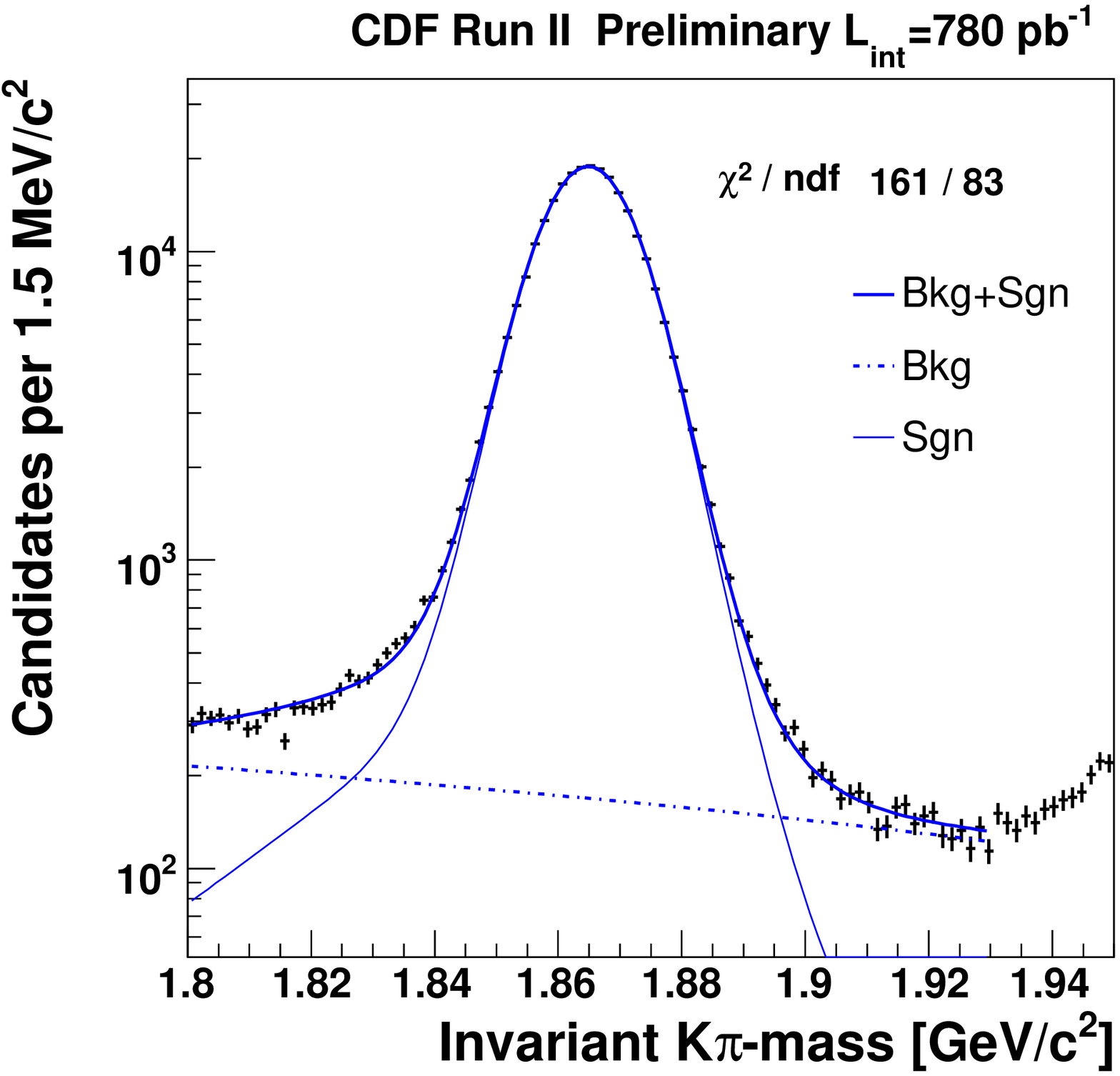,width=3.0in}
\epsfig{file=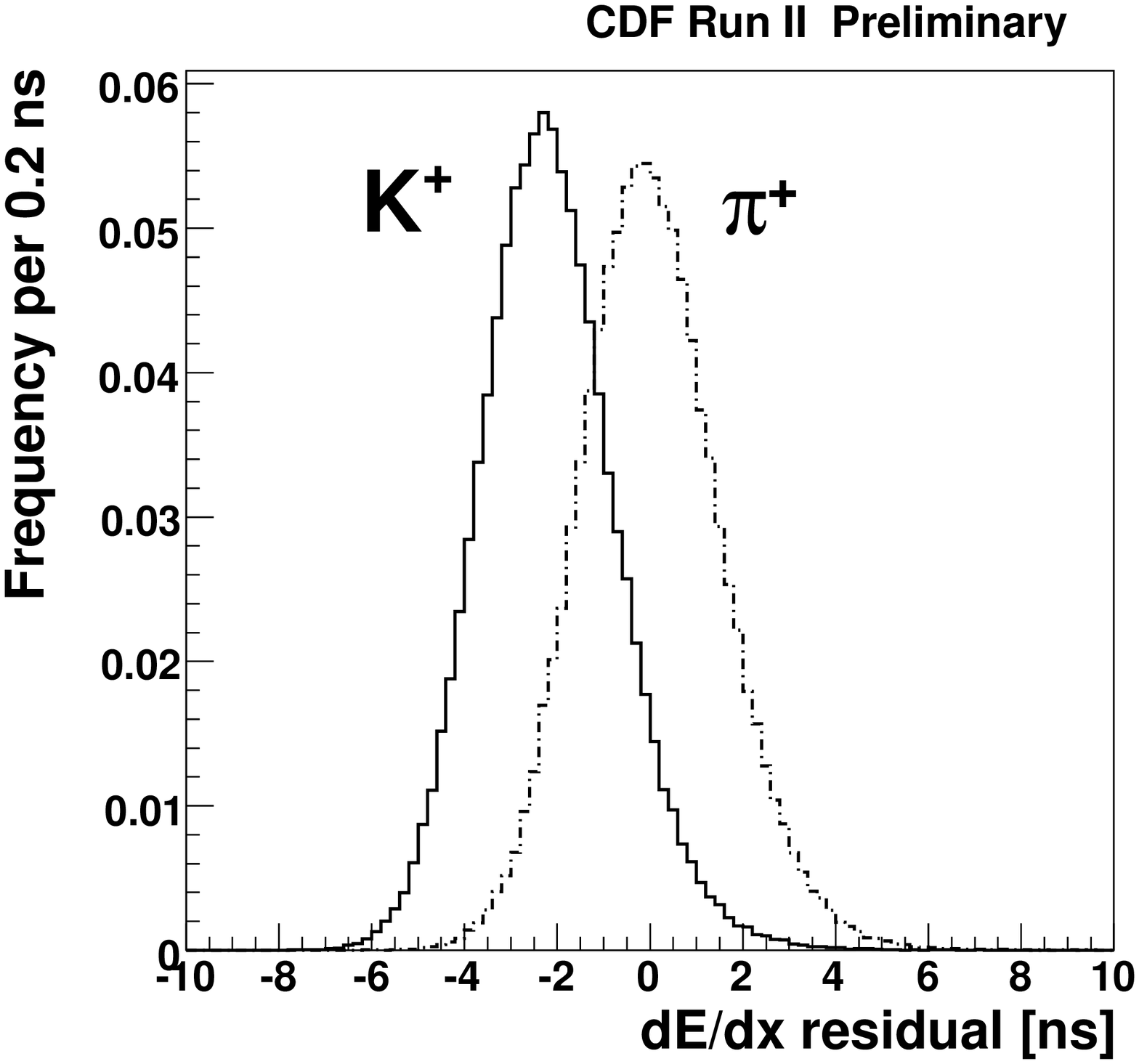,width=3.0in}
\put(-390,170){\large\bf (a)}
\put(-45,170){\large\bf (b)}
}
\caption{Tagged $D^{0} \to K^- \pi^+$ decays from $D^{*+} \to D^{0} \pi^{+}
\to [K^{-}\pi^{+}]\pi^{+}$. (a) Check of the mass line shape template
performing a 1-dimensional binned fit where the signal mass line shape is
completely fixed from the model. (b) Distribution of \dedx\ (mean COT
pulse-width) around the average pion response for calibration samples of
kaons (left) and pions (right).}
\label{fig:dstar}
\end{center}
\end{figure}

Particle identification (PID) information is summarized by a single
observable $\kappa$ for each track defined as
\begin{equation}
\kappa =\frac{\dedx - \dedx(\pi)}{\dedx(K) - \dedx(\pi)}.
\end{equation}
With the chosen observables, the likelihood contribution of the
$i^{\mathit{th}}$ event is written as:
\begin{equation}
\label{eq:likelihood}
    \mathcal{L}_i =  (1-b)\sum_{j} f_j \mathcal{L}^{\mathrm{kin}}_j  \mathcal{L}^{\mathrm{PID}}_j 
+  b \left( f_{\rm{A}} \mathcal{L}^{\mathrm{kin}}_{\mathrm{A}}
    \mathcal{L}^{\mathrm{PID}}_{\mathrm{A}}+
   (1-f_{\rm{A}}) \mathcal{L}^{\mathrm{kin}}_{\mathrm{E}}
    \mathcal{L}^{\mathrm{PID}}_{\mathrm{E}}
	\right)
\end{equation}
where:
\begin{equation}
    \label{eq:signal}\mathcal{L}_j^{\mathrm{kin}}=
    R(m_{\pi\pi}-\mathcal{M}_{j}(\alpha,p_{tot}),\alpha,p_{tot})
    P_{j}(\alpha,p_{\rm{tot}}),
\end{equation}
 \begin{equation}\label{eq:bck_A}
\mathcal{L}^{\mathrm{kin}}_{\mathrm{A}} =
           {\rm A}(m_{\pi\pi}|c_{2},m_{0}) P_{\mathrm{A}}(\alpha,p_{\rm{tot}}),
\end{equation}
\begin{equation}
\label{eq:bck_E}\mathcal{L}^{\mathrm{kin}}_{\mathrm{E}} =
           e^{c_{1} m_{\pi\pi}} P_{\mathrm{E}}(\alpha,p_{\rm{tot}}),
\end{equation}
\begin{equation}
\label{eq:PID}\mathcal{L}^{\mathrm{PID}}_{j\mathrm{(E,A)}} =
    F_{j\mathrm{(E,A)}}(\kappa_{1}, \kappa_{2},\alpha,p_{\rm{tot}}).
\end{equation}
The index `$\mathrm{A(E)}$' labels the physical (combinatorial)
background-related quantities, the index $j$ runs over the twelve
distinguishable \Bhh\ and $\Lb\to ph$ modes
(Fig.~\ref{fig:mpipi_vs_alpha}), and $f_j$ are their respective fractions,
to be determined by the fit together with the total background fraction $b$
and with the fraction of the physical (combinatorial) background
$f_{A(E)}$.  The conditional probability density
$R(m_{\pi\pi}-\mathcal{M}_{j}(\alpha,p_{tot}),\alpha,p_{tot})$ is the mass
resolution function of each mode $j$ when the correct mass is assigned to
both tracks.  In fact, the average mass $\mathcal{M}_{j}(\alpha,p_{tot})$
is the value of $m_{\pi\pi}$ obtained from Eq.~(\ref{eq:Mpipi2}) by setting
the appropriate particle masses for each decay mode $j$. Making a simple
variable change,
$R(m_{\pi\pi}-\mathcal{M}_{j}(\alpha,p_{tot}),\alpha,p_{tot})=
R(m_{j}-m_{B^0(B^{0}_{s},\Lambda^0_b)},\alpha,p_{tot})$ is obtained where
$m_{j}$ is the invariant mass computed with the correct mass assignment to
both particles for each mode $j$.  $R$ is parameterized using the detailed
detector simulation~\cite{cdfSim}. To take into account non-Gaussian tails
due to the emission of photons in the final state, CDF includes in the
simulation soft photon emission of particles in agreement with recent QED
calculations \cite{Cirigliano-Isidori}. CDF checks the quality of the mass
resolution model using about 500K \DKpi\ decays as shown in
Figure~\ref{fig:dstar}(a). The mass line-shape of the \DKpi\ peak is fitted
fixing the signal shape from the model, only allowing to vary the
background function. CDF obtains good agreement between data and
simulation. In Eq.~(\ref{eq:signal}) the nominal \Bd, \Bs\ and \Lb\ masses
as measured by CDF~\cite{CDFmasspaper} are used in order to cancel common
systematic uncertainties.  The background mass distribution is determined
in the fit by varying the parameters $c_1$, $c_2$ and $m_{0}$ in
Eq.~(\ref{eq:bck_A},\ref{eq:bck_E}).  The probability
$P_{j}(\alpha,p_{tot})$ is the joint probability distribution of $(\alpha,
p_{tot})$ and is parameterized for each mode $j$ by a product of polynomial
and exponential functions fitted to Monte Carlo samples produced by a
detailed detector simulation~\cite{cdfSim}. The background function
$P_{\mathrm{A(E)}}$ is obtained from the mass sidebands of the data.

A sample of 1.5M $D^{*+}\to D^0\pi^+ \to [K^-\pi^+]\pi^+$ decays, where the
$D^0$ decay products are identified by the charge of the $D^{*+}$ pion, was
used to calibrate the \dedx\ response over time and over the entire
tracking volume, and to determine the $F$ functions in Eq.~(\ref{eq:PID}).
Using a $>95\%$ pure $D^0$ sample, CDF obtains a $1.4\,\sigma$ separation
between kaons and pions as shown in Fig.~\ref{fig:dstar}(b), corresponding
to an uncertainty on the measured fraction of each class of particles that
is just 60\% worse than the uncertainty attainable with ideal separation.
The background term in Eq.~(\ref{eq:PID}) is similar to the signal terms,
but allows for independent pion, kaon, proton, and electron components,
which are free to vary independently. Muons are indistinguishable from
pions with the available \dedx\ resolution.

\subsubsection{Fit Results}  

CDF performs two separate fits. The first one uses the cuts optimized to
measure the direct \acpbdkpi\ and the second one is optimized to measure
\br(\BsKpi).  Significant signals are seen for the \Bdpipi, \BdKpi, and
\BsKK\ modes, previously observed by CDF~\cite{paper_bhh}. Three new rare
modes are observed for the first time: \BsKpi, \Lbppi\ and \LbpK, while no
evidence is obtained for the \Bspipi\ and \BdKK~decay channels.

To convert the yields returned from the fit into relative branching
fractions, CDF applies corrections for efficiencies of trigger and offline
selection requirements for the different decay modes.  The relative
efficiency corrections between various modes do not exceed 20\%.  Most
corrections are determined from the detailed detector
simulation~\cite{cdfSim}, with some exceptions which are measured using
data.  A momentum-averaged relative isolation efficiency between \Bs\ and
\Bd~mesons of $1.07 \pm 0.11$ is determined from fully-reconstructed
samples of $\Bs \to J/\psi\,\phi$, $\Bs \to D^{-}_{s}\pi^{+}$, $\Bd \to
J/\psi\,K^{*0}$, and $\Bd \to D^{-}\pi^{+}$.  The lower specific ionization
of kaons with respect to pions in the drift chamber is responsible for a
$\simeq 5$\% lower efficiency to reconstruct a kaon.  This effect is
measured in a sample of $D^{+}\to K^{-}\pi^{+}\pi^{+}$ decays triggered
with the two track trigger, using the unbiased third track.  The only
correction needed by the direct $CP$~asymmetries \acpbdkpi\ and \acpbskpi\
is a $\le 0.6\%$ shift due to the different probability for $K^{+}$ and
$K^{-}$ to interact with the tracker material. This correction uses a
sample of 1M prompt \DKpi\ decays reconstructed and selected with the same
criteria as the \Bhh\ decays. Assuming the Standard Model expectation
$\acpDKpi=0$, the difference between the number of reconstructed \DKpi\
decays and $\bar{D}^{0} \to K^+ \pi^-$ provides a measurement of the
detector-induced asymmetry between $K^+\pi^-$ and $K^-\pi^+$ final states.
Since CDF uses the same fit technique developed for the \Bhh\ decays, this
measurement provides also a robust check on all possible charge asymmetry
biases of the detector and \dedx\ parameterizations.

The \BsKK\ and \Bspipi\ modes require a special treatment, since they
contain a superposition of the flavour eigenstates of the \Bs~meson.  Their
time evolution might differ from the one of the flavour-specific modes if
the width difference $\Delta\Gamma_{s}$ between the \Bs~mass eigenstates is
significant.  The current result is derived under the assumption that both
modes are dominated by the short-lived \Bs\ component, that means
$\Gamma_s=\Gamma_d$, and $\Delta\Gamma_s/\Gamma_s = 0.12\pm
0.06$~\cite{Beneke:1998sy,Lenz:2004nx}.  The latter uncertainty is included
in estimating the overall systematic uncertainty.

The dominant contributions to the systematic uncertainty are as
follows. The statistical uncertainty on the isolation efficiency (\Bs\
modes), the uncertainty on the \dedx\ calibration and parameterization and
the uncertainty of the combinatorial background model.  The first one is
the larger systematics of all measurements with the meson $B^0_s$ in the
initial state except for \acpbskpi. This uncertainty is preliminary and
conservative, a significant improvement is expected for the final results.
The second one, due to \dedx, is a large systematics of all measurements,
although the parameterization of the specific ionization $\dedx$ is very
accurate.  The fit of the sample composition is very sensitive to the PID
information.  The third systematic error is due to the statistical
uncertainty of the possible combinatorial background models and it is a
dominant systematics for the observables of the rare modes.  Smaller
systematic uncertainties are assigned for the trigger efficiencies,
physical background shapes and kinematics, and the $B$~meson masses and
lifetimes.

\begin{table}[tb]
\begin{center}
{\footnotesize
\begin{tabular}{lc|lc|c}
\hline
Mode & N$_{\rm signal}$ & Quantity & Measurement & \br~[$10^{-6}$]  \\
\hline
\BdKpi         & $4045\pm84$          &  ${\cal A}_{CP}(\Bd)$         
& $-0.086\pm0.023\pm0.009$   &                            \\
\Bdpipi        & $1121\pm63$          & \BdpipisuBdKpidef\ 
& $0.259\pm0.017\pm0.016$    & $5.10\pm0.33\pm0.36$ \\
\BsKK          & $1307\pm64$          & \BsKKsuBdKpidef\   
& $0.324\pm0.019\pm0.041$    & $24.4\pm1.4\pm4.6$   \\
\hline
\BsKpi              & $230\pm34\pm16$  & \BsKpisuBdKpidef\ 
&  $0.066\pm0.010\pm0.010$ & $5.0\pm0.75\pm1.0$   \\
                    &                        & ${\cal A}_{CP}(\Bs)$
&  $0.39\pm0.15\pm0.08$    &                            \\
                    &                        & ${\cal A}_{\Gamma}(\Bs)$   
&  $-3.21\pm1.60\pm0.39$   &                            \\
\Bspipi             & $26\pm16\pm14$   &\BspipisuBdKpidef\ 
&  $0.007\pm0.004\pm0.005$ & $0.53\pm0.31\pm0.40$ \\
		    &                        &      		 
&  		                  & ($< 1.36$ @~90\%~CL)       \\
\BdKK               & $61\pm25\pm35$  & \BdKKsuBdKpidef\  
&  $0.020\pm0.008\pm0.006$ & $0.39\pm0.16\pm0.12$ \\
		    &                        &      		 
&  	                          &  ($< 0.7$ @~90\%~CL)       \\
\LbpK               & $156\pm20\pm11$  & \LbppisuLbpKdef\   
&  $0.66\pm0.14\pm0.08$    &                            \\
\Lbppi              & $110\pm18\pm16$  &                   
&                                &                            \\
\hline
\end{tabular}
}
\caption{Results on data sample optimized to measure \acpbdkpi\ (top) and
 \br(\BsKpi) (bottom).  Absolute branching fractions are normalized to the
 world-average values ${\mathcal B}(\mbox{\BdKpi}) = (19.7\pm 0.6)
 \times 10^{-6}$, $f_{s}= (10.4 \pm 1.4)\%$ and $ f_{d}= (39.8 \pm
 1.0)\%$~\cite{HFAG06}. We use ${\cal A}_{CP}(\Bd)=\ACPddef$, ${\cal
 A}_{CP}(\Bs)=\ACPsdef$ and ${\cal A}_{\Gamma}(\Bs)= \rateratiodef$.  The
 first quoted uncertainty is always statistical, the second is systematic.}
\label{tab:summary}
\end{center}
\end{table}

The measured relative branching fractions are listed in
Table~\ref{tab:summary}, where $f_{d}$ and $f_{s}$ indicate the respective
production fractions of \Bd\ and \Bs~mesons from the fragmentation of
$b$~quarks in $\bar{p}p$ collisions.  An upper limit is also quoted for
modes in which no significant signal is observed~\cite{F-C}. The absolute
branching fraction results listed are obtained by normalizing the data to
the world-average of \br(\BdKpi)~\cite{HFAG06}.

CDF reports the first observation of three new rare charmless decays
\BsKpi, \Lbppi\ and \LbpK\ with a significance respectively of $8.2
\sigma$, $6 \sigma$ and $11.5 \sigma$.  The significance includes both
statistical and systematic uncertainty. The statistical uncertainty to
evaluate the significance is estimated using several pseudo-experiments
with no contributions from rare signals.

The rate of the newly observed mode $\br(\BsKpi)=(5.0 \pm 0.75 \pm
1.0)\cdot 10^{-6}$ is in agreement with the latest theoretical expectation
\cite{zupan} which is lower than previous predictions~\cite{B-N,Yu-Li-Cai}.
CDF measures for the first time in the $B_{s}^0$~meson system the direct
$CP$~asymmetry $\acpbskpi=0.39 \pm 0.15 \pm 0.08$.  This value favors a
large $CP$~violation in \Bs~mesons, on the other hand it is also compatible
with zero. Ref.~\cite{Lipkin-BsKpi} suggests a robust test of Standard
Model expectations versus new physics comparing the direct $CP$~asymmetries
in the \BsKpi\ and \BdKpi\ decay modes.  Using HFAG input~\cite{HFAG06},
CDF measures
$\frac{\Gamma(\aBdKpi)-\Gamma(\BdKpi)}{\Gamma(\BsKpi)-\Gamma(\aBsKpi)} =
0.84 \pm 0.42 \pm 0.15$ (where $\Gamma$ is the decay width) in agreement
with the Standard Model expectation of one. Assuming that the relationship
above yields one and using as input the branching fraction \br(\BsKpi)
measured in this analysis, the world average for \acpbdkpi\ and the
\br(\BdKpi) \cite{HFAG06}, the expected value for $\acpbskpi \approx 0.37$
is estimated in agreement with the CDF measurement.

The rate of the mode $\br(\BsKK)= (24.4 \pm 1.4 \pm 4.6) \cdot 10^{-6}$ is
in agreement with the latest theoretical
expectation~\cite{matiasBsKK,matiasBsKK2} and with the previous CDF
measurement \cite{paper_bhh}.  An improved systematic uncertainty is
expected for the final analysis of the same sample.  The results for the
\Bd~meson are in agreement with world average values~\cite{HFAG06}.  The
measurement $\acpbdkpi =-0.086 \pm 0.023 \pm 0.009$ is the world's second
best measurement and the significance of the new world average
$A^{avg}_{CP}(\BdKpi)=-0.095 \pm 0.013$ moved from 6$\sigma$ to 7$\sigma$.
CDF updates the upper limits and quotes also the absolute branching
fractions of the currently unobserved annihilation-type modes: \BdKK\ and
\Bspipi.  The rate $\br(\BdKK) = (0.39 \pm 0.16 \pm 0.12)\cdot 10^{-6}$ has
the same uncertainty as the current measurements~\cite{HFAG06}, while the
\Bspipi\ upper limit (already the world's best limit~\cite{paper_bhh}) is
improved by a factor of 1.3, approaching the expectations from recent
calculations~\cite{Bspipi,Bspipi-PQCD}.  CDF also reports the first
observation of two new baryon charmless modes \Lbppi\ and \LbpK, and
measures $\LbppisuLbpK = 0.66 \pm 0.14 \pm 0.08$ in agreement with
expectations from Ref.~\cite{Mohanta}.

\section{Summary}

We review recent result on heavy quark physics focusing on Run\,II
measurements of $B$~hadron spectroscopy and decay at the Tevatron. A wealth
of new $B$~physics measurements from CDF and D\O\ has been available. These
include the spectroscopy of excited $B$~states ($B^{**}$, $B_s^{**}$) and
the observation of the \Sb~baryon.  The discussion of the decays of
$B$~hadrons and measurements of branching fractions focuses on charmless
two-body decays of \Bhh. We report several new \Bs\ and \Lb~decay channels.

\subsubsection*{Acknowledgments}

I would like to thank the organizers of this stimulating meeting for an
excellent conference. I also thank my colleagues from the CDF and
D\O~collaboration for their help in preparing this talk as well as these
proceedings. I also would like to thank Ann, Emma and Helen, a constant
source of inspiration and support, for their continuous understanding about
the life of a traveling particle physicist.  This work was supported in
part by the U.S.~Department of Energy under Grant No.~DE-FG02-91ER40682.

\end{document}

%% file: econfmacros.tex
\def\beq{\begin{equation}}
\def\eeq#1{\label{#1}\end{equation}}
\def\eeqn{\end{equation}}

\def\beqa{\begin{eqnarray}}
\def\eeqa#1{\label{#1}\end{eqnarray}}
\def\eeqan{\end{eqnarray}}

\let\bar=\overbar

\def\O{{\cal O}}

\def\Dslash{\not{\hbox{\kern-4pt $D$}}}
\def\dslash{\not{\hbox{\kern-2pt $\del$}}}

\def\msb{{\bar{\ssstyle M \kern -1pt S}}}




%% file: HQL06_proc.bbl
\newpage
\begin{thebibliography}{99}


\bibitem{bfeasi} 
N.~Ellis and A.~Kernan,
Phys.\ Rept.\  {\bf 195} (1990) 23.

\bibitem{cdf_firstB}
F.~Abe {\it et al.}  [CDF Collaboration],
Phys.\ Rev.\ Lett.\  {\bf 68} (1992) 3403.

\bibitem{myrevart}
M.~Paulini, 
Int.\ J.\ Mod.\ Phys.\  A {\bf 14} (1999) 2791
[arXiv:hep-ex/9903002].

\bibitem{cdfup}
D.~Acosta {\it et al.}  [CDF Collaboration],
Phys.\ Rev.\  D {\bf 71} (2005) 032001
[arXiv:hep-ex/0412071].

\bibitem{dup}
V.~M.~Abazov {\it et al.}  [D0 Collaboration],
Nucl.\ Instrum.\ Meth.\  A {\bf 565} (2006) 463
[arXiv:physics/0507191].

\bibitem{PDG}
W.~M.~Yao {\it et al.}  [Particle Data Group],
J.\ Phys.\ G {\bf 33} (2006) 1.

\bibitem{eichten}
E.~J.~Eichten, C.~T.~Hill and C.~Quigg,
Phys.\ Rev.\ Lett.\  {\bf 71} (1993) 4116
[arXiv:hep-ph/9308337].

\bibitem{Ebert}
D.~Ebert, V.~O.~Galkin and R.~N.~Faustov,
Phys.\ Rev.\  D {\bf 57}, 5663 (1998)
[Erratum-ibid.\  D {\bf 59}, 019902 (1999)]
[arXiv:hep-ph/9712318].

\bibitem{Isguretal}
N.~Isgur,
Phys.\ Rev.\  D {\bf 57} (1998) 4041.

M.~Di Pierro and E.~Eichten,
Phys.\ Rev.\  D {\bf 64} (2001) 114004
[arXiv:hep-ph/0104208].

\bibitem{BdsLEP_OPAL}
R.~Akers {\it et al.}  [OPAL Collaboration],
Z.\ Phys.\  C {\bf 66} (1995) 19.

\bibitem{BdsLEP}
P.~Abreu {\it et al.}  [DELPHI Collaboration],
Phys.\ Lett.\  B {\bf 345} (1995) 598.

D.~Buskulic {\it et al.}  [ALEPH Collaboration],
Z.\ Phys.\  C {\bf 69} (1996) 393.

R.~Barate {\it et al.}  [ALEPH Collaboration],
Phys.\ Lett.\  B {\bf 425} (1998) 215.

A.~A.~Affolder {\it et al.}  [CDF Collaboration],
Phys.\ Rev.\  D {\bf 64} (2001) 072002.

\bibitem{Falk:1995th}
  A.~F.~Falk and T.~Mehen,
  Phys.\ Rev.\  D {\bf 53} (1996) 231
  [arXiv:hep-ph/9507311].

\bibitem{Rosner:2006yk}
  J.~L.~Rosner,
  Phys.\ Rev.\  D {\bf 75} (2007) 013009
  [arXiv:hep-ph/0611207].

\bibitem{Stanley:1980fe}
D.~P.~Stanley and D.~Robson,
Phys.\ Rev.\ Lett.\  {\bf 45}, 235 (1980).

D.~P.~Stanley and D.~Robson,
Phys.\ Rev.\ D {\bf 21}, 3180 (1980).

J.~L.~Basdevant and S.~Boukraa,
Z.\ Phys.\  C {\bf 30} (1986) 103.

A.~Martin and J.~M.~Richard,
Phys.\ Lett.\ B {\bf 185}, 426 (1987).

W.~Y.~P.~Hwang and D.~B.~Lichtenberg,
Phys.\ Rev.\  D {\bf 35} (1987) 3526.

W.~Kwong, J.~L.~Rosner and C.~Quigg,
Ann.\ Rev.\ Nucl.\ Part.\ Sci.\  {\bf 37}, 325 (1987).

J.~G.~Korner, M.~Kramer and D.~Pirjol,
Prog.\ Part.\ Nucl.\ Phys.\  {\bf 33} (1994) 787
[arXiv:hep-ph/9406359].

K.~C.~Bowler {\it et al.}  [UKQCD Collaboration],
Phys.\ Rev.\ D {\bf 54}, 3619 (1996)
[arXiv:hep-lat/9601022].

E.~Jenkins,
Phys.\ Rev.\ D {\bf 54}, 4515 (1996)
[arXiv:hep-ph/9603449].

N.~Mathur, R.~Lewis and R.~M.~Woloshyn,
Phys.\ Rev.\ D {\bf 66}, 014502 (2002)
[arXiv:hep-ph/0203253].

C.~Albertus, J.~E.~Amaro, E.~Hernandez and J.~Nieves,
Nucl.\ Phys.\ A {\bf 740}, 333 (2004)
[arXiv:nucl-th/0311100].

D.~Ebert, R.~N.~Faustov and V.~O.~Galkin,
Phys.\ Rev.\ D {\bf 72}, 034026 (2005)
[arXiv:hep-ph/0504112].

\bibitem{Gronau:2000md}
  M.~Gronau and J.~L.~Rosner,
  Phys.\ Lett.\  B {\bf 482} (2000) 71
  [arXiv:hep-ph/0003119].

\bibitem{Lipkin-BsKpi}
  H.~J.~Lipkin,
  Phys.\ Lett.\  B {\bf 621} (2005) 126
  [arXiv:hep-ph/0503022].

\bibitem{HFAG06}
  E.~Barberio {\it et al.}  [Heavy Flavor Averaging Group (HFAG)],
  arXiv:hep-ex/0603003.

\bibitem{B-N}
  M.~Beneke and M.~Neubert,
  Nucl.\ Phys.\  B {\bf 675} (2003) 333
  [arXiv:hep-ph/0308039].

\bibitem{Bspipi}
  Y.~D.~Yang, F.~Su, G.~R.~Lu and H.~J.~Hao,
  Eur.\ Phys.\ J.\  C {\bf 44} (2005) 243
  [arXiv:hep-ph/0507326].

\bibitem{Burasetal}
  A.~J.~Buras, R.~Fleischer, S.~Recksiegel and F.~Schwab,
  Nucl.\ Phys.\  B {\bf 697} (2004) 133
  [arXiv:hep-ph/0402112].

\bibitem{gp0308063}
  G.~Punzi,
{\it In the Proceedings of PHYSTAT2003: Statistical Problems in Particle
  Physics, Astrophysics, and Cosmology, Menlo Park, California, 8-11 Sep 
2003, pp MODT002} 
  [arXiv:physics/0308063].

\bibitem{cdfSim}
  E.~Gerchtein and M.~Paulini,
{\it Proceedings of 2003 Conference for Computing in High-Energy and
  Nuclear Physics (CHEP 03), La Jolla, California, 24-28 Mar 2003, pp 
  TUMT005} 
  [arXiv:physics/0306031].

\bibitem{Cirigliano-Isidori}
  E.~Baracchini and G.~Isidori,
  Phys.\ Lett.\  B {\bf 633} (2006) 309
  [arXiv:hep-ph/0508071].

\bibitem{CDFmasspaper}
  D.~Acosta {\it et al.}  [CDF Collaboration],
  Phys.\ Rev.\ Lett.\  {\bf 96} (2006) 202001
  [arXiv:hep-ex/0508022].

\bibitem{paper_bhh}
  A.~Abulencia {\it et al.}  [CDF Collaboration],
  Phys.\ Rev.\ Lett.\  {\bf 97} (2006) 211802
  [arXiv:hep-ex/0607021].

\bibitem{Beneke:1998sy}
  M.~Beneke, G.~Buchalla, C.~Greub, A.~Lenz and U.~Nierste,
  Phys.\ Lett.\  B {\bf 459} (1999) 631
  [arXiv:hep-ph/9808385].

\bibitem{Lenz:2004nx}
  A.~Lenz,
  arXiv:hep-ph/0412007.

\bibitem{F-C}
  G.~J.~Feldman and R.~D.~Cousins,
  Phys.\ Rev.\  D {\bf 57} (1998) 3873
  [arXiv:physics/9711021].

\bibitem{zupan}
  A.~R.~Williamson and J.~Zupan,
  Phys.\ Rev.\  D {\bf 74} (2006) 014003
  [Erratum-ibid.\  D {\bf 74} (2006) 03901]
  [arXiv:hep-ph/0601214].

\bibitem{Yu-Li-Cai}
  X.~Q.~Yu, Y.~Li and C.~D.~Lu,
  Phys.\ Rev.\  D {\bf 71} (2005) 074026
  [Erratum-ibid.\  D {\bf 72} (2005) 119903]
  [arXiv:hep-ph/0501152].

\bibitem{matiasBsKK}
  S.~Descotes-Genon, J.~Matias and J.~Virto,
  Phys.\ Rev.\ Lett.\  {\bf 97} (2006) 061801
  [arXiv:hep-ph/0603239].

\bibitem{matiasBsKK2}
  S.~Baek, D.~London, J.~Matias and J.~Virto,
  JHEP {\bf 0612} (2006) 019
  [arXiv:hep-ph/0610109].

\bibitem{Bspipi-PQCD}
  Y.~Li, C.~D.~Lu, Z.~J.~Xiao and X.~Q.~Yu,
  Phys.\ Rev.\  D {\bf 70} (2004) 034009
  [arXiv:hep-ph/0404028].

\bibitem{Mohanta}
  R.~Mohanta, A.~K.~Giri and M.~P.~Khanna,
  Phys.\ Rev.\  D {\bf 63} (2001) 074001
  [arXiv:hep-ph/0006109].

\end{thebibliography}
